# Synergy-Informed Design of Platform Trials for Combination Therapies

Nan Miles Xi [1*], Man Mandy Jin [1], Lin Wang [2], Xin Huang [1]

[1] Data and Statistical Sciences, AbbVie Inc., North Chicago, IL 60064, USA

[2] Department of Statistics, Purdue University, West Lafayette, IN 47907, USA

* Correspondence: nan.xi@abbvie.com

## Abstract

Combination drug therapies hold significant promise for enhancing treatment efficacy, particularly in fields such as oncology, immunotherapy, and infectious diseases. However, designing clinical trials for these regimens poses unique statistical challenges due to multiple hypothesis testing, shared control groups, and overlapping treatment components that induce complex correlation structures. In this paper, we develop a novel statistical framework tailored for early-phase translational combination therapy trials, with a focus on platform trial designs. Our methodology introduces a generalized Dunnett's procedure that controls false positive rates by accounting for the correlations between treatment arms. Additionally, we propose strategies for power analysis and sample size optimization that leverage preclinical data to estimate effect sizes, synergy parameters, and inter-arm correlations. Simulation studies demonstrate that our approach not only controls various false positive metrics under diverse trial scenarios but also informs optimal allocation ratios to maximize power. A real-data application further illustrates the integration of translational preclinical insights into the clinical trial design process. An open-source R package is provided to support the application of our methods in practice. Overall, our framework offers statistically rigorous guidance for the design of early-phase combination therapy trials, aiming to enhance the efficiency of the bench-to-bedside transition.

# 1 Introduction

Combination therapies have gained prominence in clinical research as a strategy to improve patient outcomes by harnessing the potential synergistic effects of multiple treatments. In oncology and immunotherapy, combining agents that target different pathways or mechanisms can overcome drug resistance and produce more durable responses [1]. Such results underscore the promise of combination regimens, but they also introduce statistical challenges in clinical trial design. Evaluating a combination's efficacy typically requires comparing it not only to a control but also to its individual components, in order to demonstrate that each component contributes meaningfully to the overall effect [2]. Regulatory guidelines have explicitly stated that for approval of a combination regimen, "*it is necessary to demonstrate the contribution of effect of each monotherapy to the overall combination.*" [3] Consequently, multi-arm trial designs that include the combination and each constituent therapy have become the gold standard for combination efficacy evaluation [4]. For instance, a three-arm trial might randomize patients to receive monotherapy $A$, monotherapy $B$, or the combination $A + B$, in order to compare all relevant arms within one study. In general, randomized multi-arm trials provide far more informative evidence on combination benefits than single-arm studies of a combo, which often yield ambiguous results [5]. Controlling error rates when assessing multiple hypotheses is a classical statistical challenge, which becomes even more intricate in combination therapy trials due to complex correlation structures.

Conducting several hypothesis tests in the same trial inflates the chance of obtaining a false-positive result. This inflation is quantified by the family-wise error rate (FWER), defined as the probability of rejecting at least one true null hypothesis among the family of hypotheses [6]. Classical multiplicity correction methods exist to control this risk. The simplest is the Bonferroni procedure, which guarantees the FWER is bounded by the nominal significance level $\alpha$ but can be overly conservative. A more powerful approach is Dunnett's test, which exploits the correlation among test statistics that arises from the common control group [7]. By accounting for this dependency, Dunnett's test achieves greater power than methods assuming independent tests. However, the classical Dunnett procedure is limited to comparisons that share a single control arm; it does not directly address the complex comparison structures that arise in combination therapy trials, where different experimental arms share treatment components. Several studies have extended Dunnett's framework for multi-arm trials—such as Bretz et al. for parametric MCP-Mod [8], Urach and Posch for multi-arm group-sequential designs [9], and Robertson et al. for response-adaptive randomization [10]. Nonetheless, these methods are not specifically tailored to the correlation structures arising in combination therapy platforms that involve both a common control and overlapping treatment components. In particular, they do not provide closed-form power expressions or allocation-optimization strategies when synergy between components is present. These gaps motivate the development of our generalized Dunnett's procedure, which explicitly incorporates the full correlation matrix among combination and monotherapy arms while retaining analytical tractability to support power analysis and design optimization.

Beyond the FWER, researchers have proposed alternative error metrics to more comprehensively control false positives in multi-arm settings. One such extension is the $m$-FWER, defined as the probability of making at least $m$ false rejections among multiple hypotheses [11]. In exploratory trials, investigators may tolerate one false positive and instead control the $m$-FWER to limit the chance of multiple false discoveries [12]. Controlling these probabilities is important for preserving the scientific integrity of a study: if a trial erroneously identifies two ineffective treatments as effective, it could lead to multiple futile Phase 3 trials or unsafe combination regimens. Most standard procedures are designed to control the FWER without considering other error metrics. In this work, we primarily ensure strong control of the usual FWER at a prespecified level, while also examining the risk of multiple false positives to provide a comprehensive assessment of error control. This aligns with recent suggestions that alternative error criteria could be considered in exploratory platform trials where a small number of false leads may be acceptable [13]. In regulated confirmatory settings, though, strong control of FWER remains the gold standard, and our methods give particular attention to maintaining that control even as the trial complexity grows.

Platform trials represent a modern approach to clinical testing that can greatly enhance efficiency in drug development. A platform trial is a perpetual multi-arm trial structure in which multiple experimental treatments are evaluated within a single overarching protocol [13]. In the context of combination therapy, a platform trial might test several combinations in parallel, or incorporate a new combination arm based on emerging preclinical evidence, all under one trial infrastructure. This flexibility offers tremendous advantages for accelerating the discovery of effective combinations. However, classical notions of the Type I error may not directly apply when the number of hypotheses tested can change over the course of the trial. Moreover, decision rules in platform trials often involve interim analyses and adaptive randomization, which can further complicate error control. Recent methodological development has started to tackle these issues. Meyer et al. (2022) define a set of error rate metrics tailored to an open-entry platform evaluating combination therapies and explore how design parameters affect operating characteristics [13]. The multi-arm multi-stage (MAMS) design extends group-sequential monitoring to multi-arm trials by using generalized Dunnett-type boundary adjustments accounting for correlation at interim looks [14]. The literature also underscores that correlation induced by shared controls or overlapping components in combination therapies is a pivotal factor in platform trial statistics [11]. These innovations form a foundation, but further generalization is needed to fully accommodate the complexity of combination therapy platforms, where arms may not be independent, and prior information could guide the design.

Another aspect of the combination trial methodology is how to allocate participants across arms. In a simple multi-arm trial with a common control, it is well known that equal allocation to each arm is not necessarily optimal for statistical power [15]. In the context of combination therapies, optimal allocation questions become even more complex. If we anticipate that the combination could have a larger effect due to synergy, then one might allocate fewer patients to that arm while still maintaining high power, and allocate more to the harder-to-detect comparisons. Alternatively, one may prioritize allocation to ensure adequate power for the "weakest link" hypothesis (e.g., proving *B* is better than *A*) since the success of the combination regimen would require *B* to have some effect on its own. Our study addresses these allocation questions by deriving optimal allocation schemes under various assumptions of synergy and effect sizes. We incorporate the correlation structure of treatment arms into the allocation problem, using optimization and simulation to identify allocation strategies that maximize the power while controlling false-positive rates.

Finally, we recognize the role of preclinical data in designing combination therapy trials. Before clinical trials, extensive laboratory and animal studies are often conducted to evaluate drug interactions and potential synergy [16]. These translational data can be leveraged to improve trial design. In our framework, we incorporate preclinical evidence in two ways. First, we use it to calibrate the expected effect size of the combination and monotherapies, feeding them into sample size calculations and the choice of decision boundaries. Second, we use preclinical insights to inform priors on the synergy parameter. If mechanistic studies indicate a strong synergistic mechanism, one might prioritize detecting an interaction effect and allocate sample size accordingly. Indeed, the Clinical Trial Design Task Force in National Cancer Institute recommends using preclinical rationale to guide combination trial designs, while acknowledging the uncertainties inherent in translating animal results to humans [1]. This approach allows us to borrow strength from preclinical data when estimating correlation structures and expected outcomes, thereby potentially reducing the sample size needed to achieve the desired power.

Overall, this study develops a rigorous statistical framework for platform trials evaluating combination therapies, addressing the intertwined challenges of multiplicity, optimal design, and translational study. We propose a generalized Dunnett's procedure that accommodates the complex correlation induced by shared control groups and overlapping treatment components in combination regimens. This procedure ensures strong control of false positives at the desired level when multiple comparisons are made within a platform trial setting. We further derive optimal sample allocation strategies under drug synergy, enabling more efficient use of patient resources by allocating participants to maximize the power to detect combination effects. Our framework also integrates preclinical evidence into the design phase, using translational data to inform effect size assumptions and synergy expectations. To assist practitioners in applying our method, we have developed an open-source R package available on GitHub and provides a comprehensive set of functions to carry out all steps of our design procedure. Taken together, these innovations aim to facilitate the clinical evaluation of promising combination therapies while maintaining strict error control, thereby accelerating the discovery of effective combination therapies in a scientifically sound manner.

## 2 Methodology

In this section, we outline the statistical methodology for designing and analyzing a platform trial with combination therapies. We proceed through the components of our proposed pipeline: **(1) Trial framework and statistical model** – definition of the hypotheses, test statistics, and trial parameters in a platform trial with one combination therapy; **(2) False positive control procedure** – controlling the family-wise error rate and other multiple error rates using a generalized Dunnett's procedure adapted to our setting; **(3) Allocation ratio optimization** – optimization of allocation ratios to maximize power for detecting treatment effects; **(4) Sample size determination** – search for the minimal total sample size that achieves the target power under the optimal allocation and false positive control; **(5) Extension to platform trials with $K$ substudies** – generalization of the design approach when multiple combination therapies are embedded in one platform trial. Throughout, we assume a continuous endpoint, and the general principles apply to binary or time-to-event endpoints.

## 2.1 Trial Framework and Statistical Model

**Null and alternative hypotheses.** Consider a platform trial with three arms: standard of care as control ($A$), investigational monotherapy ($B$), and combination therapy ($A + B$) (**Figure 1A**). Let $\mu_A$, $\mu_B$, and $\mu_{AB}$ denote the true mean endpoints for patients on each regimen, where a higher value is favorable. We define two primary, two-sided null hypotheses corresponding to the two comparisons of interest:

1. **Combination therapy hypothesis**
   $H_{0,1}$: The combination therapy $A + B$ is not superior to standard-of-care $A$. Formally, $H_{0,1}$: $\mu_{AB} = \mu_A$. The alternative $H_{a,1}$ is $\mu_{AB} \neq \mu_A$, indicating that adding $B$ may result in positive benefits or adverse consequences.

2. **Monotherapy hypothesis**
   $H_{0,2}$: The monotherapy $B$ is not superior to standard-of-care $A$. Formally, $H_{0,2}$: $\mu_B = \mu_A$. The alternative $H_{a,2}$ is $\mu_B \neq \mu_A$, indicating that $B$ alone may outperform or underperform compared to $A$.

**Standardized test statistics.** Let $\bar{Y}_A, \bar{Y}_B, \bar{Y}_{AB}$ denote the sample means for arms $A$, $B$, and $A + B$, respectively. We assume that the endpoints are either normally distributed or that the sample sizes are sufficiently large (>30), and that a common variance $\sigma^2$ holds across arms. By Central Limit Theorem,

$$\bar{Y}_A \sim N\left(\mu_A, \frac{\sigma^2}{n_A}\right), \quad \bar{Y}_B \sim N\left(\mu_B, \frac{\sigma^2}{n_B}\right), \quad \bar{Y}_{AB} \sim N\left(\mu_{AB}, \frac{\sigma^2}{n_{AB}}\right)$$

where $n_A, n_B$, and $n_{AB}$ denote the sample size in arms $A$, $B$, and $A + B$, respectively. We construct the standardized test statistics:

$$Z_1 = \frac{\bar{Y}_{AB} - \bar{Y}_A}{\sqrt{Var(\bar{Y}_{AB} - \bar{Y}_A)}}, \quad Z_2 = \frac{\bar{Y}_B - \bar{Y}_A}{\sqrt{Var(\bar{Y}_B - \bar{Y}_A)}} \quad (1)$$

Under their respective null hypotheses, both $Z_1$ and $Z_2$ follow a standard normal distribution $N(0,1)$. Notably, $Z_1$ and $Z_2$ are correlated due to the shared control and overlapping treatment components. Denote the correlation between $Z_1$ and $Z_2$ by $\rho$. Under the global null hypothesis, $(Z_1, Z_2)$ follows a bivariate normal distribution $N(\mathbf{0}, \mathbf{\Sigma})$, where $\mathbf{\Sigma}$ is a correlation matrix $\begin{pmatrix} 1 & \rho \\ \rho & 1 \end{pmatrix}$.

**Correlation between test statistics.** Under the global null hypothesis, the correlation $\rho$ shapes the joint density function of $Z_1$ and $Z_2$ (**Figure 2**). When $Z_1$ and $Z_2$ are independent ($\rho = 0$), the contours are circular, whereas when they are correlated (e.g., $\rho = 0.5$), the contours become elliptical. This change in contour shape directly affects the size of the rejection region and, consequently, the false positive rate. We can show that $\rho$ is a function of the correlations between study arms and their sample sizes (see **Appendix**):

$$\rho = \frac{\frac{\rho_{AB,B}}{\sqrt{n_{AB} n_B}} - \frac{\rho_{AB,A}}{\sqrt{n_{AB} n_A}} + \frac{1}{n_A}}{\sqrt{\left(\frac{1}{n_{AB}} + \frac{1}{n_A} - 2 \frac{\rho_{AB,A}}{\sqrt{n_{AB} n_A}}\right)\left(\frac{1}{n_A} + \frac{1}{n_B}\right)}} \quad (2)$$

Here, $\rho_{AB,B}$ denotes the endpoint correlation between the combination and monotherapy arms, and $\rho_{AB,A}$ denotes the endpoint correlation between the combination and the standard-of-care control arm. Both correlations arise from the overlapping treatment components. Note that under the assumption of independent treatment arms ($\rho_{AB,B} = \rho_{AB,A} = 0$), $\rho$ simplifies to

$$\frac{1}{\sqrt{(\frac{n_A}{n_{AB}} + 1)(\frac{n_A}{n_B} + 1)}}$$

which corresponds to the classical result derived by Dunnett (1955) [7].

### 2.2 False Positive Control Procedure

**False positive metrics.** In our study, we focus on three distinct measures of false positive errors that can arise when evaluating multiple treatments simultaneously. First, the **family-wise error rate (FWER)** represents the probability of making at least one false rejection. FWER is the most commonly used metric for multiple comparisons, providing a stringent safeguard against any erroneous claim of benefit. Second, the **family multiple error rate (FMER)** captures the probability of making multiple false rejections. This metric becomes particularly relevant in multi-arm trials where multiple treatments are evaluated in parallel and may all appear promising based on limited early-stage data. Finally, the **multiple superior false positives (MSFP)** quantifies the probability of incorrectly concluding that more than one treatment is superior to control. Notably, in a two-sided test, the MSFP equals half of the FMER at the same significance level, whereas in a one-sided test, the MSFP is equivalent to the FMER.

**Figure 2** presents visual representations of these three measures under the assumption that the test statistics are normally distributed. Since controlling only the FWER can be overly conservative in early-stage screening and translational studies, minimizing both the FMER and MSFP is essential for balancing the identification of promising treatments without misallocating resources.

**Conventional approaches for controlling false positives in multiple comparisons.** We consider three classical multiple testing procedures at the same significance level $\alpha$: Bonferroni, Holm, and Dunnett's test. Bonferroni's method assigns $\alpha/2$ to each of the two comparisons, whereas Holm's method ranks the two p-values and sequentially adjusts them, first comparing the smaller p-value to $\alpha/2$, then the larger one to $\alpha$. Dunnett's test, in contrast, calculates a single critical value $c^*$ and corresponding p-value such that

$$P(|Z_1| \leq c^*, |Z_2| \leq c^*) = 1 - \alpha$$

under the global null distribution $(Z_1, Z_2) \sim N\left(0, \begin{pmatrix} 1 & \rho^* \\ \rho^* & 1 \end{pmatrix}\right)$ with

$$\rho^* = \frac{1}{\sqrt{(\frac{n_A}{n_{AB}} + 1)(\frac{n_A}{n_B} + 1)}}$$

Although Bonferroni and Holm control the FWER at the nominal level $\alpha$, they ignore correlations among arms, resulting in overly conservative adjustments and reduced power. While Dunnett's test does account for the correlation arising from a shared control, it assumes independent treatments, potentially underestimating correlation effects in combination trials. Furthermore, none of these methods adequately address the FMER or MSFP, both of which can remain unacceptably high if correlations are not properly accounted for (see **Simulation Study**). Consequently, accurately identifying and incorporating arm correlations is crucial for maintaining adequate power while controlling false positives.

**Generalized Dunnett's procedure.** To address the limitations of conventional approaches, we propose a *generalized Dunnett's procedure* that explicitly accommodates non-zero correlations between treatment arms, thereby avoiding overly conservative adjustments. Instead of $\rho^*$, we utilize the previously derived $\rho$ to specify the joint distribution of test statistics $Z_1$ and $Z_2$. This procedure considers both shared control and overlapping treatment components and extends Dunnett's framework beyond controlling only FWER to also manage FMER and MSFP, metrics that are particularly relevant in the context of combination therapy trials. To control each error rate at a target level $\alpha$ without overly conservative adjustments, we solve for a positive critical value $c^*$ under the global null joint distribution. Specifically, for controlling the FWER, we require that

$$P(|Z_1| \leq c^*, |Z_2| \leq c^*) = 1 - \alpha \tag{3}$$

ensuring that the overall probability of any false rejection remains within the acceptable limit. For the FMER, we set

$$P(|Z_1| > c^*, |Z_2| > c^*) = \alpha \tag{4}$$

Similarly, to control the MSFP, we solve for

$$P(Z_1 > c^*, Z_2 > c^*) = \alpha \tag{5}$$

Once $c^*$ is determined, the adjusted two-sided p-value rejection threshold is given by

$$2[1 - \Phi(c^*)] \tag{6}$$

where $\Phi$ denotes the cumulative distribution function of the standard normal distribution. As we demonstrate in later sections, this generalized Dunnett's procedure is flexible and can be extended to control any type of false positive error in designs comprising $K$ substudies.

### 2.3 Allocation Ratio Optimization

**Power definition.** In a platform trial with combination therapies, the allocation ratio across arms is determined to maximize test power based on the anticipated effect sizes and potential synergy. Let $\delta_{AB} = \mu_{AB} - \mu_A$ represent the true difference between the combination therapy $A + B$ and the standard-of-care $A$, and let $\delta_B = \mu_B - \mu_A$ denote the true difference between the monotherapy $B$ and $A$. Substituting these differences into the numerators of the test statistics $Z_1$ and $Z_2$ in Equation (1) and then squaring them yields the Wald noncentrality parameters [17]:

$$W_1 = \frac{\delta_{AB}^2}{\sigma_{AB-A}^2}, \qquad W_2 = \frac{\delta_B^2}{\sigma_{B-A}^2}$$

Here $\sigma_{AB-A}^2$ and $\sigma_{B-A}^2$ denote the true (population) variances of the respective mean-difference endpoints. $W_1$ and $W_2$ can be interpreted as the expected values of the squared test statistics under the alternative hypotheses. In a two-sided test, power is defined as the probability that either $|Z_1|$ or $|Z_2|$ exceeds a prespecified cutoff when the test statistics follow their respective alternative distributions. Consequently, to maximize power, the trial design should aim to maximize both $W_1$ and $W_2$ by choosing an optimal allocation ratio and leveraging the anticipated effect sizes and synergy. We demonstrate this approach in detail in the following sections.

**Synergy modeling and sample allocation ratios.** In the context of combination therapy trials, an additive effect is defined as the sum of the individual effects of the constituent monotherapies relative to the control. Under strict additivity, the expected improvement of the combination therapy would equal the sum of the improvements of each monotherapy alone. We introduce a synergy parameter $s$ to model deviations from strict additivity:

$$\delta_B = \delta, \qquad \delta_{AB} = s\delta$$

where $\delta > 0$ represents the true effect of monotherapy $B$ relative to the standard-of-care $A$. The synergy parameter $s$ captures the nature of the interaction:

- $s > 1$ indicates synergy, meaning the combination effect exceeds the additive expectation.
- $s = 1$ corresponds to additivity.
- $s < 1$ suggests antagonism, where the combination effect falls short of the additive expectation.

It is worth noting that definitions of synergy can vary across disciplines. Some interpretations are based on biological interactions or dose-response curves rather than simple additive improvement in outcomes [18]. Here, we adopt an operational definition based on deviations from the additive model on the endpoint scale, consistent with conventions used in early-phase trial design and translational research settings[*]. Although the true effect $\delta$ could theoretically be negative, thereby altering the interpretation of the synergy parameter $s$, a negative $\delta$ is not clinically relevant in the context of combination therapy discovery. In practice, it is rare for a combination therapy to be effective if one component is not [19]. Moreover, since the power analysis depends on $\delta^2$ (as shown in Equation (7)), excluding negative effect sizes does not compromise the theoretical framework.

Let $n_A, n_B$, and $n_{AB}$ denote the sample sizes for each arm, which are determined by the allocation ratios $p_A, p_B$, and $p_{AB}$ such that

$$p_A + p_B + p_{AB} = 1, \qquad 0 < p_A, p_B, p_{AB} < 1$$

with

$$n_A = p_A N, \qquad n_B = p_B N, \qquad n_{AB} = p_{AB} N$$

where $N$ is the total sample size of the trial. By expressing the Wald noncentrality parameters $W_1$ and $W_2$ in terms of the effect size, synergy parameter, and allocation ratios, we obtain (see **Appendix**):

$$W_1 = \frac{Ns^2\delta^2}{\sigma^2\left(\frac{1}{p_{AB}} + \frac{1}{p_A} - 2\frac{\rho_{AB,A}}{\sqrt{p_{AB}p_A}}\right)}, \qquad W_2 = \frac{N\delta^2}{\sigma^2\left(\frac{1}{p_A} + \frac{1}{p_B}\right)} \qquad (7)$$

---

[*] Alternative frameworks for defining synergy, such as Bliss independence and Loewe additivity, are sometimes used in preclinical and pharmacological studies. In this work, we adopt an additive model on the outcome scale for simplicity and interpretability in the clinical trial design context.

where $\sigma^2$ is the common variance in each arm. This formulation provides a framework for optimizing power through appropriate sample allocation while accounting for treatment synergy.

**Power maximization.** As previously discussed, increasing $W_1$ or $W_2$ enhances the probability of rejection under the alternative, thereby improving power. Therefore, the goal is to design a trial that maximizes both $W_1$ and $W_2$. A common strategy is to maximize the minimum of these two Wald noncentrality parameters, ensuring that neither test is underpowered [20]. This is particularly important in combination therapy studies, where regulatory agencies require evidence of efficacy for both the combination and the monotherapy. Mathematically, the optimization problem is expressed as:

$$\max_{p_A, p_B, p_{AB}} \min\{W_1, W_2\} \tag{8}$$

subject to

$$p_A + p_B + p_{AB} = 1, \quad 0 < p_A, p_B, p_{AB} < 1$$

In general, there is no closed-form solution for $p_A, p_B$, and $p_{AB}$ using analytic methods, and thus a numerical approach is required. However, in certain special cases, a closed-form solution does exist. We discuss both approaches below.

- **Numerical optimization approach**

    We re-parameterize the allocation ratios using three unconstrained real parameters $\theta_0, \theta_1$, and $\theta_2$ via the softmax function:

    $$\begin{aligned} p_A &= \frac{\exp(\theta_0)}{\exp(\theta_0) + \exp(\theta_1) + \exp(\theta_2)} \\ p_B &= \frac{\exp(\theta_1)}{\exp(\theta_0) + \exp(\theta_1) + \exp(\theta_2)} \\ p_{AB} &= \frac{\exp(\theta_2)}{\exp(\theta_0) + \exp(\theta_1) + \exp(\theta_2)} \end{aligned} \tag{9}$$

    This parameterization automatically enforces the constraints $p_A + p_B + p_{AB} = 1$ and $0 < p_{AB}, p_B, p_A < 1$. We then express the Wald noncentrality parameters $W_1$ and $W_2$ in terms of $\theta_0, \theta_1$, and $\theta_2$ as shown in Equation (9). The resulting optimization problem can be solved using the `optim` function in R programming language with a Newton-based method (see **Appendix**). Notably, the total sample size $N$, effect size $\delta$, and common variance $\sigma^2$ are constant and can be factored out of $W_1$ and $W_2$, simplifying the optimization process.

The softmax transformation in Equation (9) maps a three-component vector $(\theta_0, \theta_1, \theta_2)^T$ onto the two-dimensional probability simplex, so the parameterization is slightly redundant: adding any constant to all $\theta$ leaves $p$ unchanged. This over-parameterization has no impact on the optimization, because our objective depends only on $p$, and every point on the one-dimensional ridge of equivalent $\theta$ values yields the same unique optimum $(p_A^*, p_B^*, p_{AB}^*)$ [21]. We keep the full three-parameter form because of three practical advantages. First, any unconstrained vector can be passed to standard optimizers, and no extra constraints or projections are required. Second, softmax guarantees strictly positive arm sizes throughout the search, avoiding zero-allocation singularities in variance formulas. Third, the same programming code extends seamlessly to more than three arms.

- **Close-form solution in a special case**
  A key insight in solving this optimization problem is that, at the optimum, the two Wald noncentrality parameters $W_1$ and $W_2$ must be equal. Otherwise, the allocation ratios could be adjusted to increase the smaller one and thereby improve the overall power [22]. Equating $W_1$ and $W_2$ and canceling the common factors yields

$$\frac{s^2}{\frac{1}{p_{AB}} + \frac{1}{p_A} - 2\frac{\rho_{AB,A}}{\sqrt{p_{AB}p_A}}} = \frac{1}{\frac{1}{p_A} + \frac{1}{p_B}} \tag{10}$$

which serves as an additional constraint in the optimization process. To derive a close-form solution, we assume zero correlation between the combination therapy $A + B$ and the standard-of-care $A$ (i.e., $\rho_{AB,A} = 0$). Under this assumption, Equation (10) simplifies to

$$\frac{s^2}{\frac{1}{p_{AB}} + \frac{1}{p_A}} = \frac{1}{\frac{1}{p_A} + \frac{1}{p_B}} \tag{11}$$

With the constraints $p_{AB} + p_B + p_A = 1$ and $0 < p_{AB}, p_B, p_A < 1$, we derive a target function. By taking the derivative and setting it equal to zero, we obtain a unique solution (see **Appendix**):

$$p_A^* = \frac{\sqrt{s+1} - 1}{s}, \quad p_B^* = \frac{s + 1 - \sqrt{s+1}}{s+1}, \quad p_{AB}^* = \frac{s + 1 - \sqrt{s+1}}{s(s+1)} \tag{12}$$

Thus, under the assumption of independence between the combination therapy and the standard of care, the optimal allocation ratios that maximize power are determined solely by the synergy parameter $s$.

Several practical points are noteworthy regarding power maximization. First, the effect size $\delta$ and the synergy parameter $s$ can be estimated from preclinical or early-phase clinical data (see **Real Data Analysis**). Second, the optimal allocation does not depend on the effect sizes $\delta_B$ and $\delta_{AB}$, nor on the correlation between the combination therapy and the monotherapy $\rho_{AB,B}$. Third, the assumption underlying the closed-form solution (12), namely $\rho_{AB,A} = 0$, might not universally hold in every clinical setting. Therefore, it is essential to evaluate this correlation using data from preclinical or early-stage studies before applying the closed-form solution to trial design.

## 2.4 Sample Size Determination

Given the optimal allocation ratios $(p_A^*, p_B^*, p_{AB}^*)$ and the family-wise critical value $c^*$, the final step in the trial design is to determine the minimal total sample size $N^*$ required to achieve the target power. We determine $N^*$ numerically by combining a Monte Carlo power evaluation with a stochastic binary search procedure. The process consists of three main steps:

**Step 1: Locating an upper bound**

We begin by identifying an upper bound for the sample size. Starting from an initial small value $N_0$, we estimate power at $N = N_0$ using the Monte Carlo evaluation described in Step 3. If the estimated power is below the target, we double the current $N$ and repeat the evaluation. This process continues until the power criterion is satisfied. The first value of $N$ meeting the power requirement is stored as $N_{max}$, and the previous value is recorded as $N_{min}$.

**Step 2: Binary search**

Once the upper and lower bounds are established, we perform a binary search to find the minimal $N^*$ that satisfies the power requirement. We initialize low $= N_{min}$ and high $= N_{max}$. While low $<$ high, we compute

$$\text{mid} = \left\lfloor \frac{\text{low} + \text{high}}{2} \right\rfloor$$

and evaluate the power at $N = \text{mid}$. If the estimated power is at least the target, we update high $=$ mid; otherwise, we set low $=$ mid $+ 1$. The algorithm terminates when low $=$ high, and this value is taken as the minimal sample size $N^*$ that achieves the desired power.

**Step 3: Monte Carlo power evaluation**

For each candidate sample size $N$, we estimate power using a Monte Carlo procedure. Given the optimal allocation ratios $(p_A^*, p_B^*, p_{AB}^*)$ and inter-arm correlations $\rho_{AB,A}$ and $\rho_{AB,A}$, we compute the correlation $\rho$ using Equation (2). We then generate $n_{sim}$ realizations of the sample means $(\bar{Y}_A, \bar{Y}_B, \bar{Y}_{AB})$, each drawn from the distribution:

$$N\left( \begin{pmatrix} 0 \\ \delta \\ s\delta \end{pmatrix}, \begin{bmatrix} \frac{\sigma^2}{p_A N} & 0 & \frac{\rho_{AB,A}\sigma^2}{\sqrt{p_{AB}p_A}N} \\ 0 & \frac{\sigma^2}{p_B N} & \frac{\rho_{AB,B}\sigma^2}{\sqrt{p_{AB}p_B}N} \\ \frac{\rho_{AB,A}\sigma^2}{\sqrt{p_{AB}p_A}N} & \frac{\rho_{AB,B}\sigma^2}{\sqrt{p_{AB}p_B}N} & \frac{\sigma^2}{p_{AB}N} \end{bmatrix} \right)$$

under the alternative hypothesis (see **Appendix** for details). For each simulated vector ($\bar{Y}_A$, $\bar{Y}_B, \bar{Y}_{AB}$), we compute the test statistics ($Z_1, Z_2$) using Equation (1), and count how many times $|Z_1|$ and $|Z_2|$ exceed $c^*$. The minimum of the two empirical rejection proportions across simulations is taken as the estimated power for the given sample size $N$.

## 2.5 Extension to Platform Trials with $K$ Substudies

**Hypotheses and test statistics.** We consider a platform trial design consisting of $K$ substudies, where in each substudy, two investigational arms are compared against a common standard-of-care control arm $A$ used throughout the trial (**Figure 1B**). In each substudy $k \in \{1,2,\ldots,K\}$, one arm evaluates the combination therapy $A + B_k$ while the other assesses the monotherapy $B_k$. The true mean endpoint for the control, monotherapy, and combination therapy arms are denoted by $\mu_A$, $\mu_{B_k}$, and $\mu_{AB_k}$, respectively, with higher values indicating better patient outcomes. The sample sizes for substudy $k$ are denoted by $n_A$ for the control arm, $n_{B_k}$ for the monotherapy arm, and $n_{AB_k}$ for the combination arm.

Throughout this study, we focus on combinations involving exactly two active components. This restriction enables closed-form expressions for the optimal allocation ratios, correlation structure, and power formulas derived in Sections 2.3–2.5. Combinations involving three or more components introduce additional contrasts, higher-dimensional covariance structures, and more complex multiplicity adjustments. Extending the methodology to accommodate such higher-order combinations is an important next step and is outlined as future work in the **Discussion**.

We define two primary two-sided null hypotheses for each substudy:

1. **Combination therapy hypothesis:**
   $H_{0,k}$: The combination $A + B_k$ is not superior to standard-of-care $A$, i.e., $\mu_{AB_k} = \mu_A$. The corresponding alternative $H_{a,k}$ is $\mu_{AB_k} \neq \mu_A$, indicating that adding $B$ may yield positive or negative effects.

2. **Monotherapy hypothesis:**
   $H_{0,k+1}$: The monotherapy $B_k$ is not superior to standard-of-care $A$, i.e., $\mu_{B_k} = \mu_A$. The corresponding alternative $H_{a,k+1}$ is $\mu_{B_k} \neq \mu_A$, suggesting that $B$ alone may perform better or worse than $A$.

We assume that the endpoint measures are approximately normally distributed or that the sample sizes are sufficiently large (>30) so that the Central Limit Theorem applies. With a common variance $\sigma^2$ across arms, the sample means follow:

$$\bar{Y}_A \sim N\left(\mu_A, \frac{\sigma^2}{n_A}\right), \quad \bar{Y}_{B_k} \sim N\left(\mu_{B_k}, \frac{\sigma^2}{n_{B_k}}\right), \quad \bar{Y}_{AB_k} \sim N\left(\mu_{AB_k}, \frac{\sigma^2}{n_{AB_k}}\right)$$

For each substudy, we define two primary test statistics:

$$Z_{k,1} = \frac{\bar{Y}_{AB_k} - \bar{Y}_A}{\sqrt{Var(\bar{Y}_{AB_k} - \bar{Y}_A)}}, \quad Z_{k,2} = \frac{\bar{Y}_{B_k} - \bar{Y}_A}{\sqrt{Var(\bar{Y}_{B_k} - \bar{Y}_A)}}$$

In general, for an investigational arm $i \in \{A + B_k, B_k\}$ in substudy $k$, the test statistic is defined as

$$Z_{k,i} = \frac{\bar{Y}_{k,i} - \bar{Y}_A}{\sqrt{Var(\bar{Y}_{k,i} - \bar{Y}_A)}} \tag{13}$$

where $\bar{Y}_{k,i}$ represents the mean endpoint of the respective investigational arm. Under the global null hypotheses, each $Z_{k,i}$ follows a standard normal distribution $N(0,1)$.

**Correlation between test statistics.** Because the same control arm $A$ is shared across substudies and some treatment arms overlap in their components, test statistics are correlated both within and across substudies. Let $Z_{k,i}$ and $Z_{l,j}$ represent two test statistics from (possibly different) substudies $k$ and $l$, where $i$ and $j$ indicate either the combination or monotherapy arms. Their correlation, $Cor(Z_{k,i}, Z_{l,j})$, is given by (see **Appendix** for derivations):

$$Cor(Z_{k,i}, Z_{l,j}) = \frac{\frac{\rho_{(k,i),(l,j)}}{\sqrt{n_{k,i} n_{l,j}}} - \frac{\rho_{(k,i),A}}{\sqrt{n_{k,i} n_A}} - \frac{\rho_{(l,j),A}}{\sqrt{n_{l,j} n_A}} + \frac{1}{n_A}}{\sqrt{\left(\frac{1}{n_{k,i}} + \frac{1}{n_A} - 2\frac{\rho_{(k,i),A}}{\sqrt{n_{k,i} n_A}}\right)\left(\frac{1}{n_{l,j}} + \frac{1}{n_A} - 2\frac{\rho_{(l,j),A}}{\sqrt{n_{l,j} n_A}}\right)}} \tag{14}$$

Here, $\rho_{(k,i),(l,j)}$ denotes the endpoint correlation between arm $i$ in substudy $k$ and arm $j$ in substudy $l$, while $\rho_{(k,i),A}$ is the correlation between arm $i$ in substudy $k$ and the control arm $A$; $\rho_{(l,j),A}$ is defined similarly. The magnitude of these correlations depends on the extent of overlapping components between arms. For example, if arm $i$ in substudy $k$ and arm $j$ of substudy $l$ share common monotherapies, then $\rho_{(k,i),(l,j)} > 0$; otherwise, it is reasonable to assume $\rho_{(k,i),(l,j)} = 0$. Similarly, $n_{k,i}$ denotes the sample size of arm $i$ in substudy $k$, and $n_{l,j}$ that of arm $j$ in substudy $l$.

Since there are $2K$ test statistics across the $K$ substudies, we can compute each pairwise correlation to form a $2K \times 2K$ correlation matrix $\Sigma$. Under the global null hypothesis, the vector of sample means $(\bar{Y}_A, \bar{Y}_{B_k}, \bar{Y}_{AB_k})$ satisfies a multivariate Central Limit Theorem [23]. Since every $Z_{k,i}$ is obtained by a linear transformation of that vector, the joint distribution of all test statistics follows a multivariate normal distribution:

$$\mathbf{Z} = (Z_{1,1}, Z_{1,2}, Z_{2,1}, Z_{2,2}, \ldots, Z_{K,1}, Z_{K,2})^T \sim N(\mathbf{0}, \Sigma)$$

This distribution provides a flexible basis for controlling false positives in platform combination trials spanning $K$ substudies.

**Generalized Dunnett's procedure.** The generalized Dunnett's procedure proposed in Section 2.2 can be extended to control false positives across $K$ tests. Given the $2K$-dimensional test statistic vector $\mathbf{Z}$ and its correlation matrix $\boldsymbol{\Sigma}$, various metrics can be used to quantify false positives. Here, we demonstrate two commonly used metrics and their corresponding control strategies: the standard FWER and the $m$-FWER.

- **Standard FWER.** This metric represents the probability of making at least one false rejection among $2K$ tested hypotheses. To control the FWER at a significance level $\alpha$, we require that

$$P(\text{at least one false rejection}) = \alpha$$

This is achieved by finding a critical value $c^*$ such that

$$P\left(\bigcup_{k=1}^{K}\bigcup_{i=1}^{2}\{|Z_{k,i}| > c^*\}\right) = \alpha$$

which is equivalent to

$$P\left(\max_{k,i}|Z_{k,i}| \leq c^*\right) = 1 - \alpha$$

Once $c^*$ is determined, the adjusted two-sided p-value rejection threshold for all $2K$ tests is given by

$$2[1 - \Phi(c^*)]$$

where $\Phi$ denotes the cumulative distribution function of the standard normal distribution.

- **$m$-FWER.** In exploratory studies aimed at identifying promising treatments, it may be acceptable to tolerate up to $m$ ($m > 1$) false rejections [12]. The $m$-FWER at level $\alpha$ is defined as

$$P(\text{at least } m \text{ false rejections}) = \alpha$$

To control the $m$-FWER at level $\alpha$, we determine a critical value $c^*$ such that

$$P\left(\sum_{k=1}^{K}\sum_{i=1}^{2}\mathbb{I}\{Z_{k,i} > c^*\} \geq m\right) = \alpha$$

where $\mathbb{I}(\cdot)$ is the indicator function. The p-value is then calculated as described above.

Both control methods rely on the multivariate normal distribution of $\mathbf{Z}$ and involve integrating over the correlation structure encoded in $\boldsymbol{\Sigma}$. In practice, numerical methods are used to compute the critical values and define rejection regions that control the chosen error rate (see **Simulation Study**).

**Allocation ratio optimization.** Similar to the single-substudy setting, our objective is to find the optimal allocation ratio across all arms in a platform trial with $K$ substudies to maximize overall power. In each substudy $k$, there are three allocation ratios: $p_A$ for the common control arm $A$, $p_{B_k}$ for the monotherapy arm $B_k$, and $p_{AB_k}$ for the combination arm $A + B_k$. These allocation ratios satisfy

$$p_A + \sum_{k=1}^{K}(p_{B_k} + p_{AB_k}) = 1, \qquad 0 < p_A, p_{B_k}, p_{AB_k} < 1 \tag{15}$$

Consequently, the sample sizes for each arm in substudy $k$ are

$$n_A = p_A N, \qquad n_{B_k} = p_{B_k} N, \qquad n_{AB_k} = p_{AB_k} N$$

We assume that each monotherapy $B_k$ has a true effect $\delta_k$ over the standard-of-care $A$, defined as $\delta_k = \mu_{B_k} - \mu_A$. For the combination arm $A + B_k$, its effect is expressed as

$$\delta_{AB_k} = s_k \delta_k$$

where $s_k$ is the synergy parameter in substudy $k$. Under the alternative hypothesis, the Wald noncentrality parameters for the combination and monotherapy arms are given by (see **Appendix** for details):

$$W_{k,1} = \frac{N s_k^2 \delta_k^2}{\sigma^2 \left(\frac{1}{p_{AB_k}} + \frac{1}{p_A} - 2\frac{\rho_{AB_k,A}}{\sqrt{p_{AB_k} p_A}}\right)}, \qquad W_{k,2} = \frac{N \delta_k^2}{\sigma^2 \left(\frac{1}{p_A} + \frac{1}{p_{B_k}}\right)} \tag{16}$$

Here, $\rho_{AB_k,A}$ denotes the endpoint correlation between the combination arm in substudy $k$ and the control arm, and $\sigma^2$ is the common variance across arms. These Wald noncentrality parameters quantify how effectively each test detects a departure from the null hypothesis. To ensure that no single comparison is underpowered, we maximize the minimum of these statistics across all substudies:

$$\max_{p_A, p_{B_k}, p_{AB_k}} \min_{1 \leq k \leq K} \{W_{k,1}, W_{k,2}\}$$

Since a close-form solution is generally unavailable, we employ a numerical approach based on softmax re-parameterization. Specifically, let $\boldsymbol{\theta} = (\theta_0, \theta_1, \dots, \theta_{2K})$ be a vector of unconstrained real parameters, and define:

$$p_A = \frac{\exp(\theta_0)}{\sum_{i=0}^{2K} \exp(\theta_i)}, \qquad p_{AB_k} = \frac{\exp(\theta_{2k-1})}{\sum_{i=0}^{2K} \exp(\theta_i)}, \qquad p_{B_k} = \frac{\exp(\theta_{2k})}{\sum_{k=0}^{2K} \exp(\theta_i)}$$

for $k = 1,2,...,2K$. This re-parameterization automatically enforces the constraints specified in Equation (15). The optimization problem is then reformulated as

$$\max_{\boldsymbol{\theta}} \min_{1 \leq k \leq K} \{W_{k,1}(\boldsymbol{\theta}), W_{k,2}(\boldsymbol{\theta})\}$$

which can then be solved using standard numerical methods. During optimization, each iteration updates parameter vector $\boldsymbol{\theta}$, recalculates the allocation ratios, and evaluates the Wald noncentrality parameters. Upon convergence, this procedure yields an allocation strategy that maximizes the minimum power across all $2K$ comparisons, given the specified synergy parameters $\{s_k\}$ and effect sizes $\{\delta_k\}$ for $k = 1,2,...,K$. In practice, the values for $\{\delta_k\}$ and $\{s_k\}$ can be estimated from preclinical or early-phase clinical data (see **Real Data Analysis**).

**Sample size determination.** The overall procedure for determining the minimal total sample size $N^*$ – involving an upper bound search, binary search, and Monte Carlo power evaluation – remains the same as in the single-substudy setting described in Section 2.4. The key differences in the general $K$-substudy case are as follows:

**Step 3: Monte Carlo power evaluation (generalization)**

In the general case with $K$ substudies, we generate $n_{sim}$ Monte Carlo samples of $(2K + 1)$-dimensional vectors of sample means $(\bar{Y}_A, \bar{Y}_{B_1}, \bar{Y}_{AB_1}, ..., \bar{Y}_{B_K}, \bar{Y}_{AB_K})$, drawn from a multivariate normal distribution under the alternative hypothesis

$$N(\boldsymbol{\mu}, \boldsymbol{\Sigma})$$

where the $(2K + 1)$-dimensional mean vector $\boldsymbol{\mu} = (0, \delta_1, s_1\delta_1, ..., \delta_K, s_K\delta_K)^T$ encodes the noncentrality for each arm. The $(2K + 1) \times (2K + 1)$ correlation matrix $\boldsymbol{\Sigma}$ accounts for the correlation structure among overlapping arms, with diagonals as $\sigma^2/(p_jN)$ and off-diagonals as $\rho_{A,B_k}\sigma^2/(\sqrt{p_Ap_{B_k}}N), \rho_{AB_k,A}\sigma^2/(\sqrt{p_{AB_k}p_A}N), \rho_{AB_k,B_l}\sigma^2/(\sqrt{p_{AB_k}p_{B_l}}N)$. Each simulated vector of sample means is transformed into a vector of test statistics $\mathbf{Z} = (Z_{1,1}, Z_{1,2}, Z_{2,1}, Z_{2,2}, ..., Z_{K,1}, Z_{K,2})^T$ using Equation (13). The absolute value of each component in $\mathbf{Z}$ is compared to the common critical value $c^*$. The estimated power at $N$ is defined as the minimum empirical rejection probability across all $2K$ comparisons, ensuring sufficient power for each component of the combination trial.

## 3 Simulation Study

In this section, we present a comprehensive simulation study that demonstrates how to apply the proposed method for designing a platform trial with combination therapies. The study addresses three key objectives: **(1) Investigating how correlation structures influence different false-positive rates; (2) Comparing classical multiple testing adjustment methods with the proposed generalized Dunnett's procedure; (3) Optimizing allocation ratios and computing** **the required sample size**. For each objective, we begin by outlining the simulation setup, then describe the computational procedures, and finally discuss the insights gained from the results. For simplicity and ease of visualization, we focus on a single-substudy trial ($K = 1$), which includes one combination therapy $A + B$, one monotherapy $B$, and a shared standard-of-care control $A$. The same approach naturally extends to trials with multiple substudies ($K \geq 2$).

## 3.1 Impact of Arm Correlations on False Positives

**Parameter setup.** We manipulate key correlation parameters between arms to assess their influence on false positive rates under various error metrics. Specifically, we examine $\rho_{AB,B}$, the endpoint correlation between the combination therapy $A + B$ and the monotherapy $B$, and $\rho_{AB,A}$, the endpoint correlation between the combination arm $A + B$ and the standard-of-care control $A$. In each simulation, one correlation parameter is varied continuously from 0.05 to 0.95 in increments of 0.01, while the other is held constant at 0.3. For each combination of correlation values, we consider four distinct allocation ratios among the control, monotherapy, and combination therapy (see **Figure 3**). A nominal significance level of $\alpha = 0.05$ is maintained across all settings. We compare three error metrics: family-wise error rate (FWER), family-multiple error rate (FMER), and multiple superior false positives (MSFP).

**Simulation process.** For each combination of inter-arm correlations and allocation ratios, we first compute the corresponding correlation between test statistics $\rho$ using Equation (2). Next, we use the `mvrnorm` function from the R package `MASS` (version 7.3-65) to generate 100,000 random draws from a bivariate normal distribution with mean zero and correlation $\rho$, simulating the joint distribution of the test statistics under the global null hypothesis. For the simulated data, false-positive rates are calculated as follows: FWER – the proportion of simulations where at least one absolute test statistic exceeds the critical value of 1.96 (corresponding to $\alpha = 0.05$); FMER – the proportion of simulations where both absolute test statistics exceed 1.96; MSFP – the proportion of simulations where both test statistics themselves exceed 1.96. These metrics are then plotted against the manipulated correlations $\rho_{AB,B}$ and $\rho_{AB,A}$ for each allocation ratio to illustrate their impact (**Figures 3 and 4**). Each plot also includes the false positive rate for an independent trial design, i.e., separate tests of $A + B$ vs. $A$ and $B$ vs. $A$ without sharing control, which serves as baselines comparisons.

**Key findings.** As $\rho_{AB,B}$ increases from 0.05 to 0.95, the FWER starts below the independent-trial baseline and continues to decrease, while both FMER and MSFP increase and remain above their baselines (**Figure 3**). This pattern holds across different allocation ratios, suggesting that as the two investigational arms behave similarly, the risk of simultaneously making multiple false rejections escalates. In contrast, the effect of $\rho_{AB,B}$ depends on the allocation ratios (**Figure 4**). Specifically, when the control arm has a larger sample size, increasing $\rho_{AB,A}$ leads to a rise in FWER but a decrease in both FMER and MSFP, indicating a reduced risk of simultaneous false rejections. If the control arm is not the largest group, all metrics remain stable for $\rho_{AB,A} < 0.8$;

however, for $\rho_{AB,A} > 0.8$, FWER decreases while FMER and MSFP increase. Importantly, regardless of allocation ratios, FWER consistently stays below its baseline while FMER and MSFP are maintained above their baselines.

The simulation highlights important trade-offs among different false positive metrics. A correlation structure that reduces the chance of any false rejection (i.e., lower FWER) may simultaneously increase the risk of multiple false rejections (i.e., higher FMER or MSFP), and vice versa. In practice, the choice of false positive metric should align with the study's objectives. For example, confirmatory studies should prioritize strict control of FWER to minimize any false positives, whereas preclinical or early-phase studies – focused on identifying as many promising treatments as possible – may accept a small number of false positives but limit multiple simultaneous false rejections. Our R package provides users with the functionality to compute different false positive metrics to guide decision-making regarding these trade-offs.

### 3.2 Multiple Testing Adjustments and the Generalized Dunnett's Procedure

**Parameter setup.** We evaluate four multiple testing adjustment methods – Bonferroni, Holm, Dunnett's test, and the proposed generalized Dunnett's procedure – across a range of correlation scenarios. Specifically, we vary either $\rho_{AB,B}$ or $\rho_{AB,A}$ from 0.05 to 0.95 in increments of 0.01 while keeping the other fixed at 0.3. The overall sample allocation ratios remain identical across all arms, and the nominal significance level is set at $\alpha = 0.05$. We then compare the resulting FWER, FMER, and MSFP for each adjustment method.

**Simulation process.** For each combination of correlation parameters, we begin by computing the corresponding correlation between test statistics $\rho$ using Equation (2). We then use the `mvrnorm` function from the R package `MASS` to generate 100,000 bivariate normal samples with mean zero and correlation $\rho$, simulating the joint distribution of the test statistics under the global null hypothesis. For each simulated dataset, we determine critical values for the Bonferroni, Holm, and Dunnett's test at a nominal significance level of $\alpha = 0.05$, and apply each method to make rejection decisions as demonstrated in Section 2.2. Next, we calculate the resulting FWER, FMER, and MSFP, comparing these values to both the false positive metrics without adjustment and to the independent-trial baseline rates (**Figure 5**). Finally, we apply our generalized Dunnett's procedure to compute critical values that control FWER, FMER, and MSFP at independent-trial levels, following Equations (3), (4), and (5), respectively. This computation is implemented by first constructing the normal density function by `pmvnorm` from R package `mvtnorm` (version 1.3-3) and then using the `uniroot` function for root finding. The resulting critical values are then converted to adjusted p-values using Equation (6), with the same p-value applied uniformly to both arms for rejection decisions. These p-values are plotted against the manipulated correlations $\rho_{AB,B}$ and $\rho_{AB,A}$ in **Figure 6**.

**Key findings.** The simulation results indicate that, overall, the pattern of false positives under multiple testing adjustments is similar to that observed without adjustment (**Figure 5**). All three conventional adjustment methods maintained FWER below both the target of 0.05 and the independent-trial baseline. However, as $\rho_{AB,B}$ increases, these methods become overly conservative, driving FWER further below 0.05 and thus reducing power, either by failing to leverage the arm correlation (Bonferroni and Holm) or by misestimating it (Dunnett's test). This over-adjustment is less pronounced as $\rho_{AB,B}$ increases, given its relatively modest impact on FWER. In contrast, none of the conventional methods adequately control FMER or MSFP below their independent-trial baselines when using the nominal $\alpha = 0.05$, due to the lack of a built-in mechanism to address these error metrics.

**Figure 6** shows the p-value thresholds computed using the generalized Dunnett's procedure to control three measures of false positives under various correlation structures among arms. By design, applying these thresholds in rejection decisions guarantees control of false positives exactly at the desired targets – specifically, the FWER at 0.05, and the FMER and MSFP at levels observed in independent trials. The pattern of these p-values is roughly the inverse of the pattern of false positive metrics observed in **Figure 5**, reflecting a dynamic calibration: when arm correlations inflate false positive rates, the procedure lowers the p-value threshold to make rejection more difficult, and conversely, when false positives are deflated, it raises the p-value threshold to increase power while maintaining the false positive target. This adaptive calibration enables the generalized Dunnett's procedure to control different types of false positives at their intended targets without the over-conservatism seen in conventional methods. Our R package provides users with the functionality to compute these p-value thresholds based on user-specified false positive metrics.

### 3.3 Allocation Ratio Optimization and Sample Size Calculation

**Parameter setup.** We establish key trial parameters under the alternative hypothesis as described in Section 2.3 and simulate a comprehensive trial design pipeline. The synergy parameter $s$ is varied from 0.7 to 1.3 in increments of 0.1. Both arm correlations $\rho_{AB,A}$ and $\rho_{AB,B}$ are set to range from 0.1 to 0.7 in increments of 0.2, under the simplifying assumption that $\rho_{AB,A} = \rho_{AB,B}$. We assume an effect size of $\delta = 0.3$ and target an 80% power level. The common variance across arms is set to $\sigma^2 = 1$. Our goal is to control three error metrics – FWER at 0.05, FMER at 0.0025, and MSFP at 0.000625 – with FMER and MSFP set at the levels observed under independent trials.

**Simulation process.** For each combination of the synergy parameter and correlation settings, we first determine the optimal allocation ratios $(p_A^*, p_B^*, p_{AB}^*)$ by solving the optimization problem defined in Equation (8) via the numerical optimization approach described in Section 2.3. Under the global null hypothesis, we compute the correlation $\rho$ between the test statistics using Equation (2), based on the chosen allocation ratios and arm correlations $\rho_{AB,A}$ and $\rho_{AB,B}$. Next, we apply the generalized Dunnett's procedure – employing Equations (3) through (6) – to calculate critical values and p-value thresholds that control FWER, FMER, and MSFP at their target levels. With these thresholds fixed, we determine the minimal total sample size $N^*$ required to achieve the target power using the Monte Carlo power evaluation with the stochastic binary search procedure described in Section 2.4. The initial sample size in the binary search is set to 20, and each power estimate is based on 10,000 Monte Carlo replications. This process is repeated across all combinations of synergy and correlation settings, forming a complete pipeline to compute the required sample size and optimal allocation ratios while satisfying both false-positive control and power objectives.

**Key findings.** **Figure 7** shows the optimal allocation ratios and the corresponding sample sizes required under three false-positive control targets across various synergy and arm correlation scenarios. Notably, higher synergy parameters significantly reduce the total sample size required to achieve 80% power and shifts a larger proportion of subjects from the combination arm $A + B$ to the monotherapy arm $B$, while the allocation to the control arm $A$ remains relatively stable. This is expected because stronger synergy enhances the true effect of combination therapy, thereby reducing the sample size needed for that arm. Increased arm correlations further decrease the required total sample size, indicating that stronger correlations enhance the detectability of treatment effects. The optimal allocation ratios for the two treatment arms vary with their correlation values: the allocation to the combination arm $p_{AB}$ decreases with higher correlations, while the allocation to the monotherapy arm $p_B$ increases. Additionally, controlling FWER at 0.05 requires fewer subjects than controlling FMER or MSFP at their independent trial levels, reflecting the more stringent nature of these multi-error metrics. Overall, the proposed design strategy dynamically balances allocations across investigational arms based on synergy and arm correlations, efficiently meeting both power and false-positive control objectives. Our R package offers functionality to compute these optimal allocation ratios and required sample sizes based on user-specified synergy parameters and arm correlations, providing an integrated pipeline to support trial design.

# 4 Real Data Analysis

In this section, we demonstrate how to use the proposed methods with preclinical data to design platform combination trials. We start by describing the dataset and presenting estimates of key trial parameters. Next, we show how to control false positives using the generalized Dunnett's procedure within a fixed trial design. We then compute the optimal allocation ratios and the sample size required to achieve the power and false-positive control targets. Finally, we discuss the implications of these findings for early-stage combination trials.

**Data description.** We leverage an extensive preclinical dataset of patient-derived tumor xenograft (PDX) models to inform our trial design. Gao et al. (2015) established approximately 1,000 PDX models from tumor specimens of 277 patients, encompassing six distinct tumor types: breast, colorectal, gastric, lung, ovarian, and pancreatic cancers [24]. These models underwent comprehensive genomic profiling, including assessments of driver mutations and copy number variations, providing a robust foundation for correlating genetic alterations with therapeutic responses. In vivo drug screenings were conducted using a 1×1×1 experimental design—one mouse per model per treatment—across 62 therapeutic compounds administered both as monotherapies and combination therapies. Each PDX line was expanded into multiple mice, with a median of approximately 21 treatments evaluated per tumor. Crucially, the study produced paired endpoint data measured as the percentage of tumor size reduction for the same tumor under different treatments, enabling the estimation of treatment correlations and synergy parameters. The reproducibility and clinical translatability of these PDX models have been validated, underscoring their utility in preclinical evaluations of therapeutic strategies and in predicting clinical trial outcomes.

**Parameter estimation.** We select a subset of the PDX dataset that includes complete endpoint measures for the standard-of-care control $A$, a monotherapy $B$, and their combination $A + B$ from the same PDX model. For demonstration purposes, treatments are randomly assigned as either $A$ or $B$. For each trial, the estimated correlations $\hat{\rho}_{AB,A}$ and $\hat{\rho}_{AB,B}$ are calculated using the paired endpoint data. The effect sizes are estimated by

$$\hat{\delta}_{AB} = \frac{\bar{Y}_{AB} - \bar{Y}_A}{\hat{\sigma}_{AB,A}}, \qquad \hat{\delta}_B = \frac{\bar{Y}_B - \bar{Y}_A}{\hat{\sigma}_{A,B}}$$

where $\bar{Y}_i$ is the mean endpoint in arm $i$. $\hat{\sigma}_{AB,A}$ and $\hat{\sigma}_{A,B}$ are pooled sample standard deviations calculated using

$$\hat{\sigma}_{AB,A} = \sqrt{\frac{(n_{AB} - 1)\hat{\sigma}_{AB}^2 + (n_A - 1)\hat{\sigma}_A^2}{n_{AB} + n_A - 2}}, \qquad \hat{\sigma}_{A,B} = \sqrt{\frac{(n_A - 1)\hat{\sigma}_A^2 + (n_B - 1)\hat{\sigma}_B^2}{n_A + n_B - 2}}$$

with $n_i$ and $\hat{\sigma}_i$ representing the sample size and sample standard deviation in arm $i$, respectively. We standardize the effect sizes by dividing by the pooled standard deviations so that the resulting estimates have unit variance ($\sigma^2 = 1$). This normalization is also consistent with the common-variance structure used in the Monte Carlo power evaluation described in Section 2.4. The synergy parameter is then estimated by

$$\hat{s} = \frac{\hat{\delta}_{AB}}{\hat{\delta}_B}$$

Results with $\hat{\delta}_{AB} < 0$ or $\hat{\delta}_B < 0$ indicate a harmful treatment effect, which is not clinically relevant in the context of combination therapy discovery and is therefore omitted from subsequent trial designs. This screening yields six synthetic combination trials. The estimates, along with the corresponding drug names for $A$ and $B$, are summarized in **Table 1**.

**False positive control.** Using the estimated correlations $\hat{\rho}_{AB,A}$ and $\hat{\rho}_{AB,B}$, we compute the correlation between the test statistics for each synthetic trial by Equation (2) and denote it as $\hat{\rho}$.

The sample sizes $n_A$, $n_B$ and $n_{AB}$ are set equal to the actual number of observations for each arm in the dataset. Next, we employ the simulation method described in Section 3.1 to compute FWER, FMER, and MSFP at a nominal significance level of $\alpha = 0.05$. This step demonstrates the behavior of these false-positive metrics in the absence of any multiple testing adjustments. Subsequently, we apply the proposed generalized Dunnett's procedure to compute the corresponding p-value thresholds required to control these metrics at predetermined targets (i.e., FWER at 0.05, FMER at 0.0025, and MSFP at 0.000625) in each trial. It is important to note that we treat the sample sizes for the different arms as fixed and compute the p-value rejection thresholds accordingly. In other words, our primary goal is to ensure that the false-positive rates remain controlled despite multiple testing, given that the trial design is already determined. **Table 1** summaries the estimated correlation $\hat{\rho}$, the false positive rates without adjustment, and the p-value thresholds after adjustment.

**Allocation ratio and sample size calculation.** Different from the previous section, which focused on controlling false positives given a fixed trial design, this section presents a complete pipeline for designing platform combination trials from scratch using PDX data. Leveraging the full set of estimated parameters, $(\hat{\rho}_{AB,A}, \hat{\rho}_{AB,B}, \hat{\delta}_{AB}, \hat{\delta}_B, \hat{s})$, we calculate the optimal allocation ratios $(p_A^*, p_B^*, p_{AB}^*)$ and the minimal total sample size $N^*$ required to achieve 80% power, while ensuring that FWER, FMER, and MSFP remain below their respective control targets. **Table 2** provides a summary of the computed p-value thresholds, optimal allocation ratios, and sample sizes for each synthetic trial. Consistent with the simulation results presented in Section 3.3, our approach dynamically allocates more subjects to arms with smaller effect sizes to enhance detectability and balance power across comparisons. Notably, the estimated synergy parameters are generally high, which leads to a large value of $\hat{\delta}_{AB}$ and, consequently, a smaller allocation for the combination arm $p_{AB}^*$. Furthermore, the total sample size $N^*$ is inversely related to the effect size, and it varies considerably across the synthetic trials. This variation reflects both the wide range of estimated effect sizes and the method's ability to adapt to different effect sizes in order to achieve the desired level of detectability.

It is important to note that, unlike the simulation results illustrated in **Figure 7** – where the required sample size for controlling FMER or MSFP was always larger than that for controlling FWER – the sample size required in our real data analysis does not consistently follow this pattern. This discrepancy is due to differences in parameter settings: (1) The simulation assumed a synergy parameter ranging from 0.7 to 1.3, whereas the estimated synergy in the PDX data ranges from 1.2 to 18.4; (2) The simulation is simplified by setting $\hat{\rho}_{AB,A} = \hat{\rho}_{AB,B}$, an assumption that does not hold in the PDX estimations. These differences underscore the complex impact of arm correlations on false-positive metrics, as demonstrated in **Figures 3** and **4** and Equation (2). Consequently, the simulation and real data results are complementary rather than contradictory.

**Implications for early-stage platform combination trial design.** This case study illustrates how preclinical PDX data can be quantitatively integrated into the design of early-phase platform combination trials. In first-in-human studies, the relationships among investigational drug effects are often unknown; therefore, estimating these relationships from PDX models is critical for robust statistical planning. For example, if the combination therapy and the corresponding monotherapy exhibit highly correlated activity, more conservative error thresholds may be necessary to mitigate the risk of multiple false signals. Conversely, a lower correlation may justify a more aggressive exploration of combination benefits. Furthermore, our approach to optimizing allocation ratios and determining total sample sizes offers a practical framework for refining trial design in settings where patient availability is limited. The results suggest that intelligently allocating patients based on estimated treatment effects from preclinical studies can enhance the detection of synergy without compromising error control for the individual treatments. This strategy not only fulfills regulatory requirements by evaluating both the combination and its components but also can be extended to large-scale confirmatory Phase 3 trials. In such trials, key parameters estimated from early-stage human studies can inform the computation of optimal allocation ratios, required sample sizes, and adjusted p-value thresholds tailored to specific false-positive control and power targets..

**Software package.** We have implemented the complete pipeline of our proposed method in an open-source R package available on our GitHub repository. The package accepts as input the number of substudies, arm correlations, effect size, synergy parameter, false-positive control target, and power target, and it outputs the optimal allocation ratios, required total sample size, and corresponding p-value thresholds. Designed to accommodate large-scale platform combination trials using preclinical or early-stage data, this tool serves as a critical resource in translational research. A detailed tutorial is provided in the GitHub repository.

## 5 Discussion

This study introduces a comprehensive statistical framework to improve the design and analysis of platform trials for combination therapies. We develop a generalized Dunnett's procedure to control various types of false positives in multi-arm trials where treatment arms share components. This extends the classical Dunnett's test to more complex comparison structures with overlapping therapies. We also propose a power-optimized sample allocation strategy that leverages anticipated drug synergy. Rather than assuming equal allocation to each arm, our approach allocates patients across the combination and monotherapy arms dynamically given the expected effect sizes. In addition, we develop a design pipeline to determine the minimal total sample size required to achieve a target power while controlling false-positive rates. This pipeline allows researchers to calibrate trial size and decision criteria to meet prespecified power and error control thresholds under various correlation and effect size scenarios. Finally, our framework demonstrates the integration of preclinical data into the trial design by incorporating information from translational models. These contributions use biologically informed parameters to optimize trial design, employ a correlation-aware multiple comparison procedure to maintain statistical rigor, and ensure adequate power through efficient resource allocation.

The proposed framework offers several notable strengths with practical implications for early-phase translational research. Methodologically, it provides strong error control in the complex setting of combination trials. Our joint testing approach allows all relevant comparisons to be assessed within a single trial while rigorously controlling false positives. This unified analysis yields more informative evidence on combination efficacy than separate trials would, and it conforms to regulatory guidance that combination approvals should show the contribution of each agent. Another strength is the improvement in statistical power and efficiency achieved through the optimized allocation strategy. From a practical standpoint, this translates to smaller required sample sizes for achieving the same power, or conversely greater power for a given sample size, compared to naive equal-allocation designs. In resource-limited early-phase trials, such efficiency means faster completion and fewer patients exposed to potentially inferior treatments. Importantly, our framework is well-suited to translational and early-phase settings. The incorporation of preclinical data is a strength in contexts where human data are sparse but robust preclinical experiments exist. This preclinical-to-clinical linkage can make early-phase trials more predictive, thereby increasing the likelihood that a positive trial result will translate into a true clinical benefit. We anticipate it will be especially useful for Phase 1/2 platform trials in oncology or other fields where combinations are tested before proceeding to larger confirmatory studies.

This work builds upon and connects to several strands of existing statistical methodology. First, it contributes to the literature on multiple comparisons in multi-arm trials. Our generalized Dunnett procedure is rooted in the idea introduced by Dunnett (1955) [7], but extends it to scenarios where arms share active components. This can be viewed within the closed-testing or partitioning principle for multiple comparisons, where known correlations are utilized to adjust critical values more efficiently than assuming independence [25,26]. Our approach complements the guidance by Howard et al. (2018) on multi-arm trials with common controls [11], offering a concrete procedure to implement those principles when treatment arms are not independent. Our design framework also connects to multi-arm multi-stage (MAMS) and platform trial methodologies [13,15,27]. The key difference is that our current work is focused on fixed-design platforms with combination treatments and synergy considerations, whereas some others incorporate adaptive elements or Bayesian decision criteria. Our optimized allocation can be seen as a counterpart to adaptive randomization, achieved here through a priori calculation rather than on-the-fly adaptation. Integrating preclinical data into trial planning is another novel connection, resonating with the idea of co-clinical trials [28]. Our work connects statistical design with translational science, and complements prior efforts where preclinical studies were conducted to improve eventual clinical trial success [29]. Overall, our method stands at the intersection of multiple-testing procedures, multi-arm trial design, and translational data integration. We believe this synthesis will stimulate further methodological research and encourage more crosstalk between these previously separate areas.

Our framework has certain limitations that warrant discussion. In developing the methodology, we assumed a continuous outcome measure and applied normal theory for analytical tractability. This is appropriate for endpoints like tumor size change or biomarker levels, especially in translational research where continuous efficacy measures are common. However, many clinical trials use binary or time-to-event endpoints. While we expect that a similar multiple-comparison strategy could be applied, the lack of a current binary or survival extension is a limitation. In practice, if our method were applied to such endpoints, one would likely rely on simulation to verify error control, or consider a transformation of the outcome to approximate normality. Another limitation is that our trial design is non-adaptive in its current form. This fixed design does not incorporate interim looks or adaptive modifications, such as early stopping for efficacy/futility or dropping inferior arms mid-trial [30]. The complexity of multiple comparisons with overlapping arms makes adaptive control of error rates challenging, and we opted to first establish methods for the fixed design. A further limitation involves the reliance on preclinical data and the accuracy of the translational assumptions. If the assumed correlation structure or synergy effect from preclinical experiments is misspecified, the optimized allocation or calculated sample size may not yield the intended power in the actual trial. Users of the framework should therefore treat preclinical-informed parameters as informed guesses and consider robust design strategies (e.g. enrolling a reserve sample if results differ from expectations). Despite these limitations, we believe they are addressable and do not detract from the core contributions.

There are several natural extensions and avenues for future work building on this framework. A first priority is to generalize the methodology to binary or time-to-event endpoints. Extending our error-control procedure and power optimization to, say, a logistic regression or Cox model setting would broaden the applicability of the framework. This likely entails deriving or simulating multivariate test statistic distributions under those models, or using alternative metrics like the family-wise error rate in terms of hazard ratios [31]. Preliminary work could investigate whether a large-sample normal approximation of the test statistics still provides accurate error control in these cases, or if more complex copula or permutation approaches are needed [32]. Another extension is to incorporate adaptive and sequential trial features into our platform design. In practice, platform trials often allow for interim analyses where ineffective arms can be dropped early or new arms added as new therapies become available [33]. Integrating our combination-testing approach with group-sequential methods would allow early stopping for efficacy or futility on one or more arms while maintaining control of overall error rates [9]. Similarly, one could explore adaptive randomization procedures that update allocation ratios based on accruing data, effectively combining our optimal initial allocation with ongoing learning [10]. Such adaptations could further improve efficiency and ethical aspects by focusing patient assignments on the most promising treatments. Finally, we plan to explore applications of the framework beyond the current scope. This includes a systematic sensitivity analysis when the assumed correlation structure and synergy are misspecified. It may also involve extending to higher-order combinations, e.g., trials that evaluate triplet therapies where overlaps are even more complex. Another practical extension is to consider covariate-adjusted versions of the tests to improve power [34,35]. While this work is motivated by oncology, where combination therapies are especially prevalent, the proposed framework is broadly applicable to other fields such as immunology, virology, and neurology, where combination treatments are also commonly explored [36–38]. These efforts will further enhance the flexibility and impact of the proposed approach.

Reproducibility and transparency have been central considerations in this work. To promote accessibility and support verification of our results, we have developed an open-source R package and made it available via a public GitHub repository. This package implements all components of the framework, providing user-friendly functions to carry out the design calculations and simulations described above. In addition, we have included vignettes and documentation illustrating how to reproduce the simulation studies and numerical results presented in this paper. We welcome feedback, issue reports, and community contributions to the codebase, in keeping with the open science principles. We are committed to maintaining this resource as an up-to-date toolkit for robust combination trial design.

## Conflicts of Interest

Nan Miles Xi, Man Mandy Jin, and Xin Huang are full-time employees of AbbVie and may hold stock or stock options in the company. AbbVie had no role in the design, analysis, interpretation, or decision to publish this study. All other authors declare no conflicts of interest.

## Data Availability Statement

All materials needed to reproduce the results of this study are openly accessible.

**R package** combodesign – source code and documentation are available from https://github.com/xnnba1984/combodesign.

**Analysis code** – contained in the companion repository https://github.com/ xnnba1984/combodesign-analysis.

**Patient-derived xenograft (PDX) dataset** used in **Real Data Analysis** is publicly available and downloaded from https://www.nature.com/articles/nm.3954#Sec28.

All code is released under the MIT license. Data are redistributed under the original CC-BY 4.0 terms of the source repository.

## Funding

Author Lin Wang was supported by the National Science Foundation (DMS-2413741) and the Central Indiana Corporate Partnership AnalytiXIN Initiative.

# Appendix

## 1 Correlation between Test Statistics

In this section, we derive Equation (2) from Section 2.1 to express the correlation between test statistics in terms of the correlations among treatment arms and the sample sizes in each arm for a single-substudy platform combination trial. Recall that the test statistics are defined as

$$Z_1 = \frac{\bar{Y}_{AB} - \bar{Y}_A}{\sqrt{Var(\bar{Y}_{AB} - \bar{Y}_A)}}, \qquad Z_2 = \frac{\bar{Y}_B - \bar{Y}_A}{\sqrt{Var(\bar{Y}_B - \bar{Y}_A)}}$$

Under the global null hypothesis, both $Z_1$ and $Z_2$ follow a standard normal distribution $N(0,1)$. Their correlation $\rho$ is given by

$$\rho = \frac{Cov(Z_1, Z_2)}{\sqrt{Var(Z_1)Var(Z_2)}} = \frac{Cov(\bar{Y}_{AB} - \bar{Y}_A, \bar{Y}_B - \bar{Y}_A)}{\sqrt{Var(\bar{Y}_{AB} - \bar{Y}_A)Var(\bar{Y}_B - \bar{Y}_A)}}$$

By simple algebra, the numerator can be written as

$$Cov(\bar{Y}_{AB} - \bar{Y}_A, \bar{Y}_B - \bar{Y}_A) = C_{AB,B} - C_{AB,A} - C_{A,B} + V_A$$

where $C_{i,j} = Cov(\bar{Y}_i, \bar{Y}_j)$ and $V_i = Var(\bar{Y}_i)$. Similarly, the two variances in the denominator are

$$Var(\bar{Y}_{AB} - \bar{Y}_A) = V_{AB} + V_A - 2C_{AB,A}, \qquad Var(\bar{Y}_B - \bar{Y}_A) = V_A + V_B - 2C_{A,B}$$

Thus, the correlation between the test statistics becomes

$$\rho = \frac{C_{AB,B} - C_{AB,A} - C_{A,B} + V_A}{\sqrt{(V_{AB} + V_A - 2C_{AB,A})(V_A + V_B - 2C_{A,B})}} \tag{17}$$

Assuming a common variance $\sigma^2$ across arms, and noting that the standard deviation of a mean is given by $\sigma/\sqrt{n}$, we have

$$C_{AB,B} = \bar{\rho}_{AB,B} \frac{\sigma^2}{\sqrt{n_{AB}n_B}}, \qquad C_{AB,A} = \bar{\rho}_{AB,A} \frac{\sigma^2}{\sqrt{n_{AB}n_A}}, \qquad C_{A,B} = \bar{\rho}_{A,B} \frac{\sigma^2}{\sqrt{n_A n_B}}$$

$$V_A = \frac{\sigma^2}{n_A}, \qquad V_B = \frac{\sigma^2}{n_B}, \qquad V_{AB} = \frac{\sigma^2}{n_{AB}}$$

where $\bar{\rho}_{ij}$ is the correlation between $\bar{Y}_i$ and $\bar{Y}_j$. Substituting these into Equation (17) yields

$$\rho = \frac{\frac{\bar{\rho}_{AB,B}}{\sqrt{n_{AB}n_B}} - \frac{\bar{\rho}_{AB,A}}{\sqrt{n_{AB}n_A}} - \frac{\bar{\rho}_{A,B}}{\sqrt{n_A n_B}} + \frac{1}{n_A}}{\sqrt{\left(\frac{1}{n_{AB}} + \frac{1}{n_A} - 2\frac{\bar{\rho}_{AB,A}}{\sqrt{n_{AB}n_A}}\right)\left(\frac{1}{n_A} + \frac{1}{n_B} - 2\frac{\bar{\rho}_{A,B}}{\sqrt{n_A n_B}}\right)}} \tag{18}$$

Next, we show that under the independent and identically distributed (i.i.d.) assumption, the correlation between the means of two random variables equals the correlation between the variables themselves; that is, $\bar{\rho}_{ij} = \rho_{ij}$. Let $U$ and $V$ denote the endpoint random variables from two arms. Suppose we have i.i.d. pairs $(U_1, V_1), (U_2, V_2), \ldots, (U_n, V_n)$ drawn from their joint distribution. Although we cannot observe $U_i$ and $U_j$ simultaneously for the same individual, they are well defined within the potential outcome framework [39].

The covariance between the sample means is

$$Cov(\bar{U}, \bar{V}) = Cov\left(\frac{1}{n}\sum_{i=1}^{n} U_i, \frac{1}{n}\sum_{j=1}^{n} V_j\right) = \frac{1}{n^2}\sum_{i=1}^{n}\sum_{j=1}^{n} Cov(U_i, V_j)$$

Under the i.i.d. assumption, $U_i$ and $V_j$ are independent for $i \neq j$, so the cross terms vanish:

$$Cov(\bar{U}, \bar{V}) = \frac{1}{n^2}\sum_{i=1}^{n} Cov(U_i, V_i)$$

Let $\rho_{U,V}$ denote the correlation between random variables $U$ and $V$, and let $\bar{\rho}_{U,V}$ denote the correlation between their sample means $\bar{U}$ and $\bar{V}$. Also, let $\sigma_U$ and $\sigma_V$ be the standard deviations of $U$ and $V$, respectively. Since each pair $(U_i, V_i)$ has the same covariance $\rho_{U,V}\sigma_U\sigma_V$ for all $i$, we have

$$cov(\bar{U}, \bar{V}) = \frac{1}{n^2}\sum_{i=1}^{n} \rho_{U,V}\sigma_U\sigma_V = \frac{\rho_{U,V}\sigma_U\sigma_V}{n}$$

Since $Var(\bar{U}) = \sigma_U^2/n$ and $Var(\bar{V}) = \sigma_V^2/n$, the correlation between the means is

$$\bar{\rho}_{U,V} = \frac{Cov(\bar{U}, \bar{V})}{\sqrt{Var(\bar{U})Var(\bar{V})}} = \frac{\frac{\rho_{U,V}\sigma_U\sigma_V}{n}}{\sqrt{\frac{\sigma_U^2}{n}\frac{\sigma_V^2}{n}}} = \rho_{U,V}$$

That is, under the i.i.d. assumption, the correlation between the means equals the correlation between the individual observations. Accordingly, in our notation, we have:

$$\bar{\rho}_{AB,B} = \rho_{AB,B}, \qquad \bar{\rho}_{AB,A} = \rho_{AB,A}, \qquad \bar{\rho}_{A,B} = \rho_{A,B} \tag{19}$$

Substituting these into Equation (18) yields

$$\rho = \frac{\dfrac{\rho_{AB,B}}{\sqrt{n_{AB}n_B}} - \dfrac{\rho_{AB,A}}{\sqrt{n_{AB}n_A}} - \dfrac{\rho_{A,B}}{\sqrt{n_A n_B}} + \dfrac{1}{n_A}}{\sqrt{\left(\dfrac{1}{n_{AB}} + \dfrac{1}{n_A} - 2\dfrac{\rho_{AB,A}}{\sqrt{n_{AB}n_A}}\right)\left(\dfrac{1}{n_A} + \dfrac{1}{n_B} - 2\dfrac{\rho_{A,B}}{\sqrt{n_A n_B}}\right)}} \qquad (20)$$

In practice, investigational monotherapy $B$ is often selected to operate via a different mechanism than control $A$ to reduce cross-resistance in combination therapy [40]. Consequently, it is reasonable to assume that the endpoints for $B$ and $A$ are independent, i.e., $\rho_{A,B} = 0$. Under this assumption, Equation (20) simplifies to

$$\rho = \frac{\dfrac{\rho_{AB,B}}{\sqrt{n_{AB}n_B}} - \dfrac{\rho_{AB,A}}{\sqrt{n_{AB}n_A}} + \dfrac{1}{n_A}}{\sqrt{\left(\dfrac{1}{n_{AB}} + \dfrac{1}{n_A} - 2\dfrac{\rho_{AB,A}}{\sqrt{n_{AB}n_A}}\right)\left(\dfrac{1}{n_A} + \dfrac{1}{n_B}\right)}}$$

which is exactly Equation (2).

## 2 Wald Noncentrality Parameters

In this section, we derive Equation (7) to show that the Wald noncentrality parameters can be expressed as functions of the synergy parameter, the effect size, and the allocation ratios. Using the notation introduced in Section 2.3, the Wald noncentrality parameters are defined as

$$W_1 = \frac{\delta_{AB}^2}{\sigma_{AB-A}^2}, \qquad W_2 = \frac{\delta_B^2}{\sigma_{B-A}^2} \qquad (21)$$

Assuming a synergy parameter $s$ and an effect size $\delta$, the numerators are given by

$$\delta_B^2 = \delta^2, \qquad \delta_{AB}^2 = s^2 \delta^2$$

Replacing the population variance by sample variance, the denominators can be expressed as follows:

$$\sigma_{AB-A}^2 = Var(\bar{Y}_{AB} - \bar{Y}_A) = V_{AB} + V_A - 2C_{AB,A} = \frac{\sigma^2}{n_{AB}} + \frac{\sigma^2}{n_A} - 2\bar{\rho}_{AB,A}\frac{\sigma^2}{\sqrt{n_{AB}n_A}}$$

$$\sigma_{B-A}^2 = Var(\bar{Y}_B - \bar{Y}_A) = V_A + V_B - 2C_{A,B} = \frac{\sigma^2}{n_A} + \frac{\sigma^2}{n_B} - 2\bar{\rho}_{A,B}\frac{\sigma^2}{\sqrt{n_A n_B}}$$

where $C_{i,j} = Cov(\bar{Y}_i, \bar{Y}_j)$ and $V_i = Var(\bar{Y}_i)$. Given the sample size for each arm is the product of the total sample size and allocation ratio, and further assuming $\bar{\rho}_{A,B} = 0$, we obtain

$$Var(\bar{Y}_{AB} - \bar{Y}_A) = \frac{\sigma^2}{N}\left(\frac{1}{p_{AB}} + \frac{1}{p_A} - 2\bar{\rho}_{AB,A}\frac{1}{\sqrt{p_{AB}p_A}}\right)$$

$$Var(\bar{Y}_B - \bar{Y}_A) = \frac{\sigma^2}{N}\left(\frac{1}{p_A} + \frac{1}{p_B}\right)$$

Substituting these into the Equation (21) and incorporating the results from Equation (19) yields

$$W_1 = \frac{s^2\delta^2}{\frac{\sigma^2}{N}\left(\frac{1}{p_{AB}} + \frac{1}{p_A} - 2\rho_{AB,A}\frac{1}{\sqrt{p_{AB}p_A}}\right)} = \frac{Ns^2\delta^2}{\sigma^2\left(\frac{1}{p_{AB}} + \frac{1}{p_A} - 2\frac{\rho_{AB,A}}{\sqrt{p_{AB}p_A}}\right)}$$

$$W_2 = \frac{\delta^2}{\frac{\sigma^2}{N}\left(\frac{1}{p_A} + \frac{1}{p_B}\right)} = \frac{N\delta^2}{\sigma^2\left(\frac{1}{p_A} + \frac{1}{p_B}\right)}$$

which match Equation (7) exactly.

## 3 Numerical Approaches to Find the Optimal Allocation Ratio

In this section, we demonstrate the numerical method for solving the max-min optimization problem presented in Equation (8) and outline the key parameters involved. We use the `optim` function in R programming language (version 4.4.3) to determine the optimal allocation ratios that maximize power, given the synergy parameter, correlations between arms, and the softmax reparameterization introduced in Equation (9). Since `optim` minimizes by default, we set the objective function as $-\min(W_1^*, W_2^*)$, where

$$W_1^* = \frac{s^2}{\frac{1}{p_{AB}} + \frac{1}{p_A} - 2\frac{\rho_{AB,A}}{\sqrt{p_{AB}p_A}}}, \quad W_2^* = \frac{1}{\frac{1}{p_B} + \frac{1}{p_A}}$$

In this formulation, the allocation ratios $p_A, p_B,$ and $p_{AB}$ are reparametrized via the parameter $\theta$ defined in Equation (9). Note that the total sample size $N$, the effect size $\delta$, and the common variance $\sigma^2$ are omitted from the objective function since they are constant factors that do not affect the optimization process. We initialize the free parameter vector $\boldsymbol{\theta}$ at zero without constraints. To efficiently solve this unconstrained optimization problem, we utilize the BFGS method, a quasi-Newton algorithm that leverages gradient information to iteratively approximate the Hessian matrix [41].

## 4 Close-Form Solution for the Optimal Allocation Ratios

In this section, we analytically derive Equation (12) to solve the max-min optimization problem presented in Equation (8). Our goal is to find the allocation ratios $p_A^*, p_B^*$, and $p_{AB}^*$ that maximize $\min\{W_1, W_2\}$, where

$$W_1 = \frac{Ns^2\delta^2}{\sigma^2\left(\frac{1}{p_{AB}} + \frac{1}{p_A} - 2\frac{\rho_{AB,A}}{\sqrt{p_{AB}p_A}}\right)}, \quad W_2 = \frac{N\delta^2}{\sigma^2\left(\frac{1}{p_A} + \frac{1}{p_B}\right)}$$

This optimization problem is subject to the constraint

$$p_A + p_B + p_{AB} = 1, \quad 0 < p_A, p_B, p_{AB} < 1$$

Since constants $N$, $\delta$, and $\sigma^2$ do not affect the optimization, we define the simplified functions

$$W_1^* = \frac{s^2}{\frac{1}{p_{AB}} + \frac{1}{p_A}}, \quad W_2^* = \frac{1}{\frac{1}{p_A} + \frac{1}{p_B}}$$

Maximizing $\min\{W_1, W_2\}$ is equivalent to maximizing $\min\{W_1^*, W_2^*\}$. For a max-min optimization problem, the optimum occurs when $W_1^* = W_2^*$. Therefore, we set

$$\frac{s^2}{\frac{1}{p_{AB}} + \frac{1}{p_A}} = \frac{1}{\frac{1}{p_A} + \frac{1}{p_B}}$$

Using the constraint $p_B = 1 - p_{AB} - p_A$, let $x = p_{AB}$ and $y = p_A$ so that $p_B = 1 - x - y$. Substituting these into the equality $W_1^* = W_2^*$ and simplifying them yields the quadratic equation

$$y^2 + (2x - 1)y + (1 - s^2)x^2 + (s^2 - 1)x = 0$$

Solving for $y$ with the quadratic formula gives two solutions:

$$y_1 = 1 - (s + 1)x, \quad y_2 = (s - 1)x$$

We discard $y_2$ since it is infeasible when $s < 1$. Thus, we select

$$y = 1 - (s + 1)x$$

which implies

$$p_A = 1 - (s + 1)x, \quad p_B = sx, \quad p_{AB} = x \tag{22}$$

Substituting these expressions into $W_1^*$ and $W_2^*$ at the optimum gives

$$W_1^* = W_2^* = \frac{s^2x[1 - (s + 1)x]}{1 - sx}$$

which depends solely on $x$. Define

$$f(x) = \frac{x[1 - (s+1)x]}{1 - sx}$$

and differentiate $f(x)$ with respect to $x$, we obtain

$$f'(x) = \frac{s(s+1)x^2 - 2(s+1)x + 1}{(1 - sx)^2}$$

Setting $f'(x) = 0$ yields the critical point

$$x^* = \frac{s + 1 - \sqrt{s+1}}{s(s+1)}$$

To verify that $x^*$ corresponds to a maximum, we examine the second derivative. Rewrite $f'(x)$ as

$$f'(x) = \frac{g(x)}{h(x)}$$

with

$$g(x) = s(s+1)x^2 - 2(s+1)x + 1, \qquad h(x) = (1 - sx)^2$$

The second derivative of $f(x)$ is

$$f''(x) = \frac{g'(x)h(x) - g(x)h'(x)}{[h(x)]^2}$$

At the critical point $x^*$, $g(x^*) = 0$, so the second derivative simplifies to

$$f''(x^*) = \frac{g'(x^*)h(x^*)}{[h(x^*)]^2}$$

Since $h(x^*) > 0$, the sign of $f''(x^*)$ is determined by $g'(x^*)$. Differentiating $g(x)$ yields

$$g'(x) = 2s(s+1)x - 2(s+1)$$

Substituting $x^* = \frac{s+1-\sqrt{s+1}}{s(s+1)}$ into $g'(x)$ gives

$$g'(x^*) = 2s(s+1) \cdot \frac{s+1-\sqrt{s+1}}{s(s+1)} - 2(s+1) = -2\sqrt{s+1} < 0$$

Therefore, $f''(x^*) < 0$, confirming that $x^*$ is indeed a global maximum. Finally, substituting $x^*$ back into Equation (22), we obtain the optimal allocation ratios:

$$p_A^* = \frac{\sqrt{s+1}-1}{s}, \qquad p_B^* = \frac{s+1-\sqrt{s+1}}{s+1}, \qquad p_{AB}^* = \frac{s+1-\sqrt{s+1}}{s(s+1)}$$

This result corresponds exactly to Equation (12), achieving the maximum power.

## 5 Distribution of Arm-Level Means Used in Monte Carlo Power Evaluation

In this section, we show that the vector of sample means $(\bar{Y}_A, \bar{Y}_B, \bar{Y}_{AB})$ follows a multivariate normal distribution under the alternative hypothesis, as used in the Monte Carlo procedure described in Section 2.4. Under the alternative hypothesis, the expected values of the sample means are:

$$E(\bar{Y}_A) = 0, \qquad E(\bar{Y}_B) = \delta, \qquad E(\bar{Y}_{AB}) = s\delta$$

The variances of the sample means are obtained using the fact that the variance of the mean is the mean of the variances:

$$Var(\bar{Y}_A) = \frac{\sigma^2}{n_A}, \qquad Var(\bar{Y}_B) = \frac{\sigma^2}{n_B}, \qquad Var(\bar{Y}_{AB}) = \frac{\sigma^2}{n_{AB}}$$

Substituting the sample sizes based on allocation ratios

$$n_A = p_A N, \qquad n_B = p_B N, \qquad n_{AB} = p_{AB} N$$

we have

$$Var(\bar{Y}_A) = \frac{\sigma^2}{p_A N}, \qquad Var(\bar{Y}_B) = \frac{\sigma^2}{p_B N}, \qquad Var(\bar{Y}_{AB}) = \frac{\sigma^2}{p_{AB} N}$$

For the covariances, we assume that the control and monotherapy arms are independent, i.e., $\rho_{A,B} = 0$. Then we have:

$$Cov(\bar{Y}_A, \bar{Y}_B) = 0$$

$$Cov(\bar{Y}_{AB}, \bar{Y}_A) = \rho_{AB,A}\sqrt{Var(\bar{Y}_{AB})Var(\bar{Y}_A)} = \frac{\rho_{AB,A}\sigma^2}{\sqrt{p_{AB}p_A}N}$$

$$Cov(\bar{Y}_{AB}, \bar{Y}_B) = \rho_{AB,B}\sqrt{Var(\bar{Y}_{AB})Var(\bar{Y}_B)} = \frac{\rho_{AB,B}\sigma^2}{\sqrt{p_{AB}p_B}N}$$

Because the arm-level observations are independent across arms and i.i.d. within arm with finite second moments, the multivariate Central Limit Theorem indicates that $(\bar{Y}_A, \bar{Y}_B, \bar{Y}_{AB})$ follows a multivariate normal distribution under alternative:

$$N\left(\begin{pmatrix}0\\\delta\\s\delta\end{pmatrix},\begin{bmatrix}\dfrac{\sigma^2}{p_A N} & 0 & \dfrac{\rho_{AB,A}\sigma^2}{\sqrt{p_{AB}p_A}N}\\ 0 & \dfrac{\sigma^2}{p_B N} & \dfrac{\rho_{AB,B}\sigma^2}{\sqrt{p_{AB}p_B}N}\\ \dfrac{\rho_{AB,A}\sigma^2}{\sqrt{p_{AB}p_A}N} & \dfrac{\rho_{AB,B}\sigma^2}{\sqrt{p_{AB}p_B}N} & \dfrac{\sigma^2}{p_{AB}N}\end{bmatrix}\right)$$

This distribution is used to generate Monte Carlo samples of test statistics in the power evaluation procedure described in Step 3 of Section 2.4. The above result generalizes naturally to the case of a platform combination trial with $K$ substudies, as presented in Section 2.5.

## 6 Correlation between Test Statistics in $K$-substudies

In this section, we derive Equation (14), which generalizes the correlation formula from a single substudy (Equation (2)) to the case of $K$ substudies. For an investigational arm $i \in \{A+B_k, B_k\}$ in substudy $k$, the test statistics is defined as

$$Z_{k,i} = \frac{\bar{Y}_{k,i} - \bar{Y}_A}{\sqrt{Var(\bar{Y}_{k,i} - \bar{Y}_A)}}$$

where $\bar{Y}_{k,i}$ represents the mean endpoint for that arm. Under the global null hypothesis, $Z_{k,i}$ follows a standard normal distribution $N(0,1)$. Because each test statistic is standardized to have a variance of 1, the correlation between any two test statistics $Z_{k,i}$ and $Z_{l,j}$ is given by

$$Cor(Z_{k,i}, Z_{l,j}) = \frac{Cov(Z_{k,i}, Z_{l,j})}{\sqrt{Var(Z_{k,i})Var(Z_{l,j})}} = Cov(Z_{k,i}, Z_{l,j})$$

By definition,

$$Cov(Z_{k,i}, Z_{l,j}) = \frac{Cov(\bar{Y}_{k,i} - \bar{Y}_A, \bar{Y}_{l,j} - \bar{Y}_A)}{\sqrt{Var(\bar{Y}_{k,i} - \bar{Y}_A)Var(\bar{Y}_{l,j} - \bar{Y}_A)}}$$

Expanding the numerator yields

$$Cov(\bar{Y}_{k,i} - \bar{Y}_A, \bar{Y}_{l,j} - \bar{Y}_A) = Cov(\bar{Y}_{k,i}, \bar{Y}_{l,j}) - Cov(\bar{Y}_{k,i}, \bar{Y}_A) - Cov(\bar{Y}_A, \bar{Y}_{l,j}) + Var(\bar{Y}_A)$$

Assuming a common variance $\sigma^2$ across arms and noting that the variance of a mean is $\sigma^2/n$, we have

$$Cov(\bar{Y}_{k,i}, \bar{Y}_{l,j}) = \bar{\rho}_{(k,i),(l,j)} \frac{\sigma^2}{\sqrt{n_{k,i} n_{l,j}}}, \qquad Cov(\bar{Y}_{k,i}, \bar{Y}_A) = \bar{\rho}_{(k,i),A} \frac{\sigma^2}{\sqrt{n_{k,i} n_A}}$$

$$Cov(\bar{Y}_A, \bar{Y}_{l,j}) = \bar{\rho}_{A,(l,j)} \frac{\sigma^2}{\sqrt{n_A n_{l,j}}}, \qquad Var(\bar{Y}_A) = \frac{\sigma^2}{n_A}$$

Thus, the numerator becomes

$$\sigma^2 \left( \frac{\bar{\rho}_{(k,i),(l,j)}}{\sqrt{n_{k,i} n_{l,j}}} - \frac{\bar{\rho}_{(k,i),A}}{\sqrt{n_{k,i} n_A}} - \frac{\bar{\rho}_{A,(l,j)}}{\sqrt{n_A n_{l,j}}} + \frac{1}{n_A} \right)$$

Similarly, the denominators are

$$\sqrt{Var(\bar{Y}_{k,i} - \bar{Y}_A)} = \sqrt{Var(\bar{Y}_{k,i}) + Var(\bar{Y}_A) - 2Cov(\bar{Y}_{k,i}, \bar{Y}_A)} = \sigma \sqrt{\left( \frac{1}{n_{k,i}} + \frac{1}{n_A} - 2 \frac{\bar{\rho}_{(k,i),A}}{\sqrt{n_{k,i} n_A}} \right)}$$

$$\sqrt{Var(\bar{Y}_{l,j} - \bar{Y}_A)} = \sqrt{Var(\bar{Y}_{l,j}) + Var(\bar{Y}_A) - 2Cov(\bar{Y}_{l,j}, \bar{Y}_A)} = \sigma \sqrt{\left( \frac{1}{n_{l,j}} + \frac{1}{n_A} - 2 \frac{\bar{\rho}_{(l,j),A}}{\sqrt{n_{l,j} n_A}} \right)}$$

Combining these expressions and incorporating the results from Equation (19), the correlation between $Z_{k,i}$ and $Z_{l,j}$ is

$$Cor(Z_{k,i}, Z_{l,j}) = \frac{\frac{\rho_{(k,i),(l,j)}}{\sqrt{n_{k,i} n_{l,j}}} - \frac{\rho_{(k,i),A}}{\sqrt{n_{k,i} n_A}} - \frac{\rho_{A,(l,j)}}{\sqrt{n_A n_{l,j}}} + \frac{1}{n_A}}{\sqrt{\left( \frac{1}{n_{k,i}} + \frac{1}{n_A} - 2 \frac{\rho_{(k,i),A}}{\sqrt{n_{k,i} n_A}} \right) \left( \frac{1}{n_{l,j}} + \frac{1}{n_A} - 2 \frac{\rho_{(l,j),A}}{\sqrt{n_{l,j} n_A}} \right)}}$$

which matches Equation (14) exactly. This final expression generalizes the correlation of test statistics to any two arms, whether they belong to the same substudy ($k = l$) or to different substudies ($k \neq l$).

## 7 Wald Noncentrality Parameters in *K*-Substudies

In this section, we derive Equation (16) to show that the Wald noncentrality parameters for a platform combination trial with $K$ substudies can be expressed as functions of the synergy parameter, the effect size, and the allocation ratios. Using the notation introduced in Section 2.5, the Wald noncentrality parameters for the $k$th substudy are defined as

$$W_{k,1} = \frac{\delta_{AB_k}^2}{\sigma_{AB_k-A}^2}, \qquad W_{k,2} = \frac{\delta_{B_k}^2}{\sigma_{B_k-A}^2}$$

Assuming a synergy parameter $s_k$ and an effect size $\delta_k$ for the $k$th substudy, the numerators are given by

$$\delta_{AB_k}^2 = s_k^2 \delta_k^2, \qquad \delta_{B_k}^2 = \delta_k^2,$$

Using the same technique in the single-substudy case, the denominators can be expressed as follows:

$$\sigma_{AB_k-A}^2 = Var(\bar{Y}_{AB_k} - \bar{Y}_A) = \frac{\sigma^2}{n_{AB_k}} + \frac{\sigma^2}{n_{A_k}} - 2\rho_{AB_k,A} \frac{\sigma^2}{\sqrt{n_{AB_k} n_A}}$$

$$\sigma_{B_k-A}^2 = Var(\bar{Y}_{B_k} - \bar{Y}_A) = \frac{\sigma^2}{n_{A_k}} + \frac{\sigma^2}{n_{B_k}} - 2\rho_{A,B_k} \frac{\sigma^2}{\sqrt{n_A n_{B_k}}}$$

Given the sample size for each arm is the product of the total sample size $N$ and the corresponding allocation ratio, and further assuming $\rho_{A,B_k} = 0$, we obtain

$$\sigma_{AB_k-A}^2 = \frac{\sigma^2}{N}\left(\frac{1}{p_{AB_k}} + \frac{1}{p_A} - 2\rho_{AB_k,A} \frac{1}{\sqrt{p_{AB_k} p_A}}\right)$$

$$\sigma_{B_k-A}^2 = \frac{\sigma^2}{N}\left(\frac{1}{p_{AB_k}} + \frac{1}{p_A}\right)$$

Substituting these into the expressions for $W_{k,1}$ and $W_{k,2}$ yields

$$W_{k,1} = \frac{N s_k^2 \delta_k^2}{\sigma^2 \left(\frac{1}{p_{AB_k}} + \frac{1}{p_A} - 2\frac{\rho_{AB_k,A}}{\sqrt{p_{AB_k} p_A}}\right)}, \qquad W_{k,2} = \frac{N \delta_k^2}{\sigma^2 \left(\frac{1}{p_A} + \frac{1}{p_{B_k}}\right)}$$

which are exactly Equation (16).

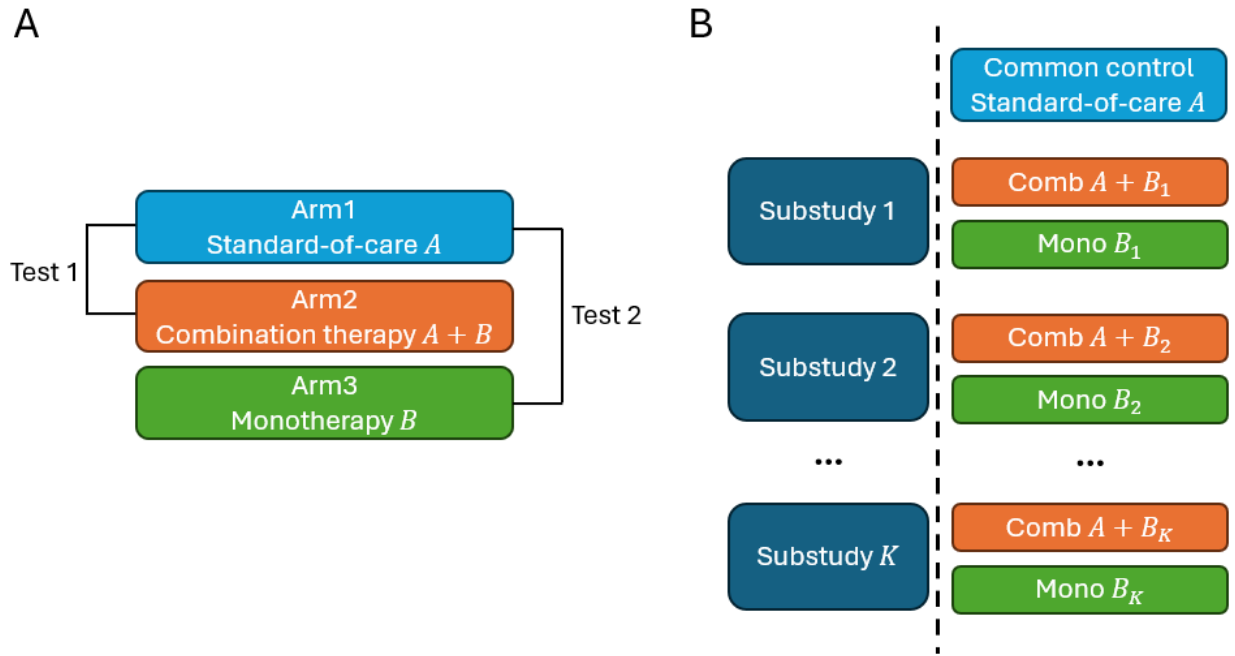

**Figure 1. Platform trial with combination therapies. A. Simplified trial design with one combination therapy.** Test 1 evaluates the efficacy of the combination therapy, while Test 2 assesses the efficacy of the monotherapy. **B. Generalized trial design with *K* substudies.** Each substudy contains one monotherapy and its corresponding combination therapy, each evaluated by its own Test 1 and Test 2. The standard-of-care *A* serves as the common control across all arms.

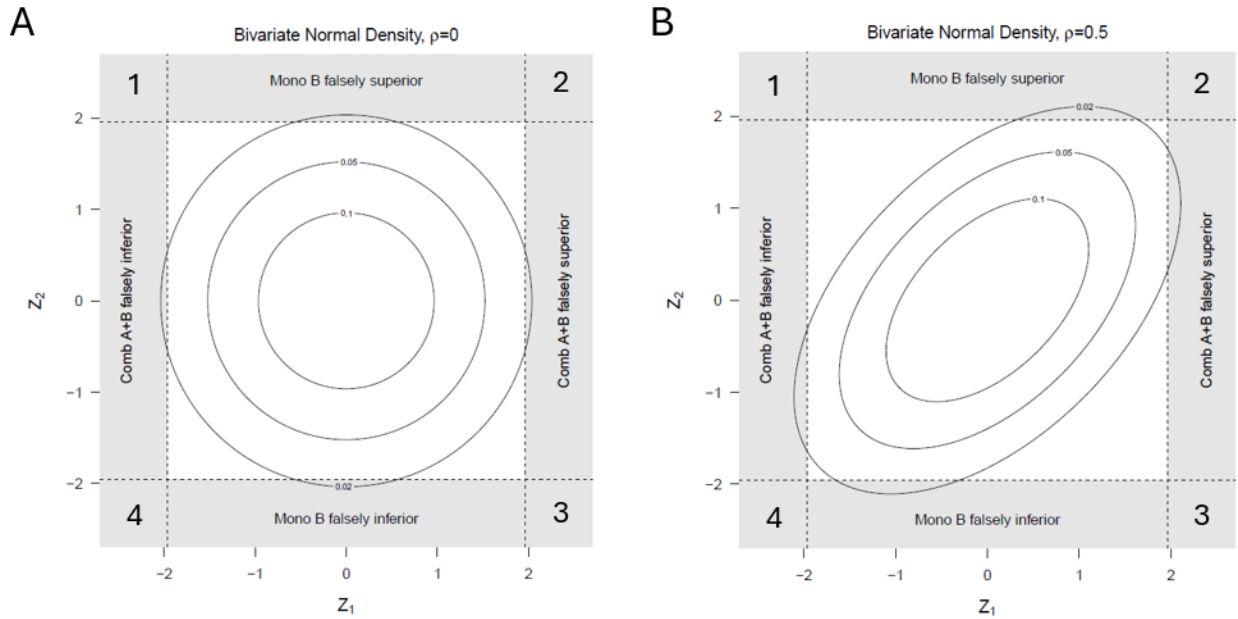

**Figure 2. Joint density functions of test statistics $Z_1$ and $Z_2$ under null hypothesis with correlation $\rho = 0$ (panel A) and $\rho = 0.5$ (panel B).** The family-wise error rate (FWER) is defined as the sum of the four shaded edge regions minus the area of regions 1+2+3+4. The family multiple error rate (FMER) is the area of regions 1+2+3+4. The multiple superior false positives (MSFP) is the area of region 2. The significance level is set at $\alpha = 0.05$. This visualization is adapted from Howard et.al. [11].

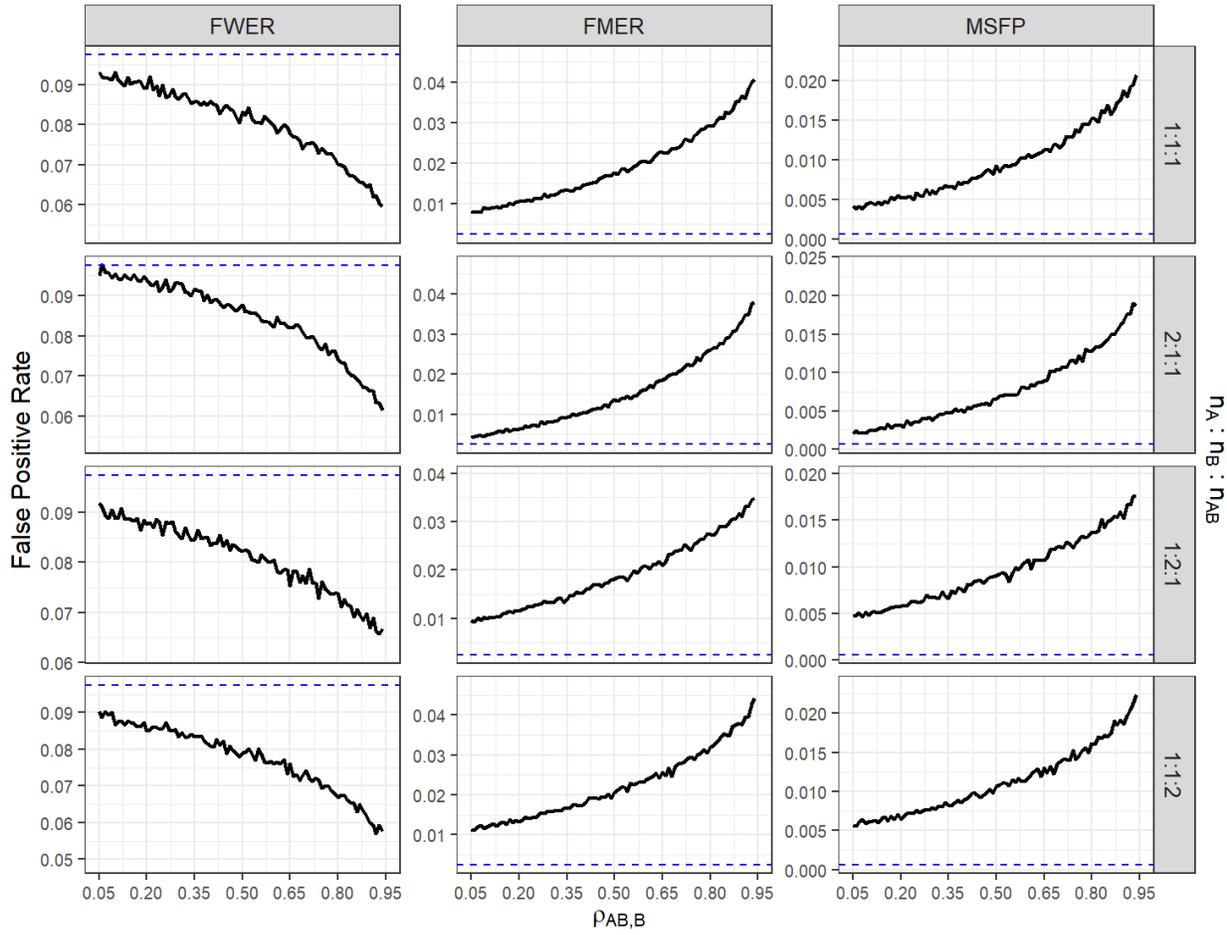

**Figure 3. False positive metrics as functions of $\rho_{AB,B}$ across different allocation ratios.** Each row corresponds to one allocation ratio, and each column shows a different metric. The black curves depict simulated false positive rates for $\rho_{AB,B}$ varying from 0.05 to 0.95. The blue dashed lines represent baseline rates under an independent-trial design (FWER=0.0975; FMER=0.0025; MSFP=0.000625). All results are generated under the global null hypothesis at a significance level of $\alpha = 0.05$.

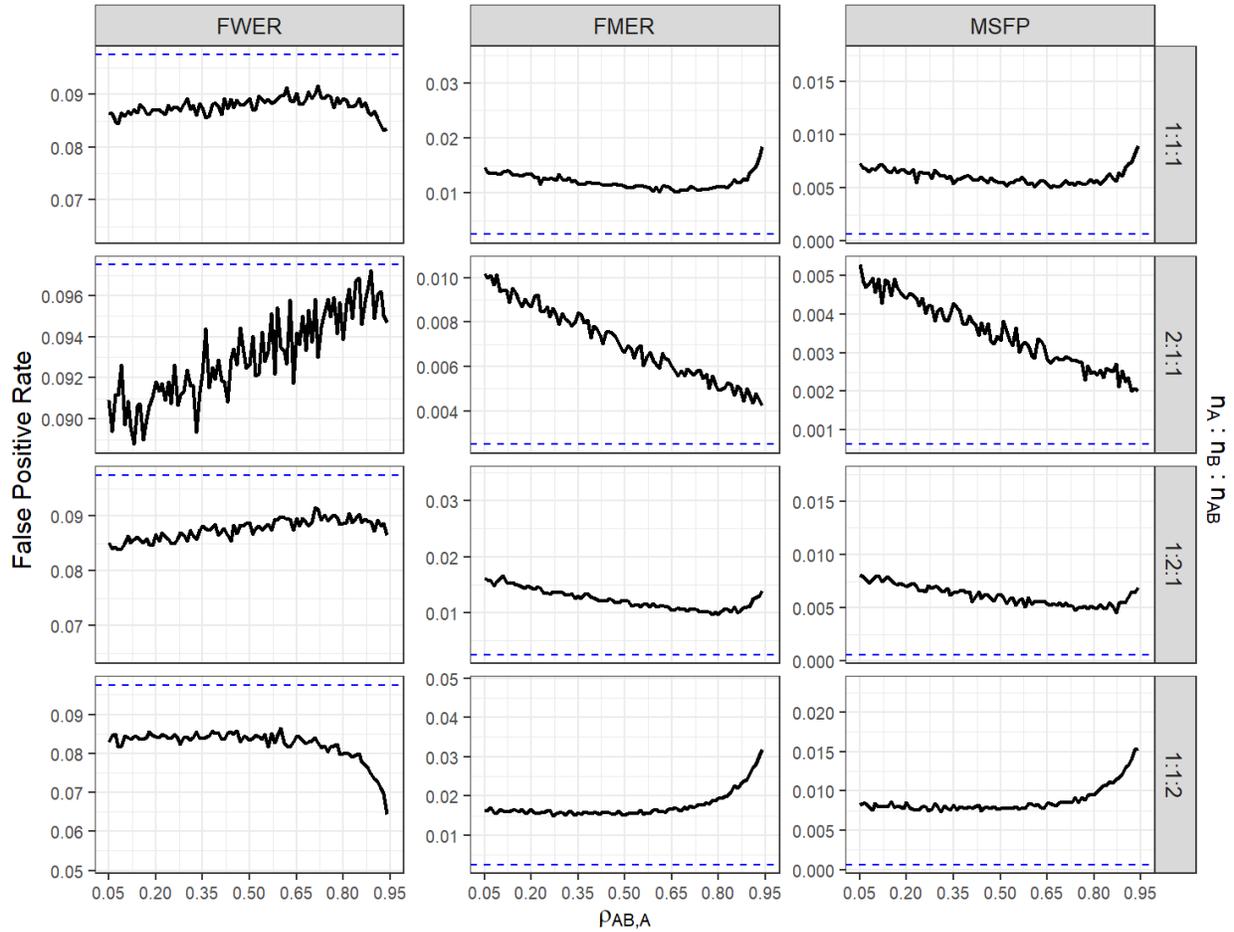

**Figure 4. False positive metrics as functions of $\rho_{AB,A}$ across different allocation ratios.** Each row corresponds to one allocation ratio, and each column shows a different metric. The black curves depict simulated false positive rates for $\rho_{AB,A}$ varying from 0.05 to 0.95. The blue dashed lines represent baseline rates under an independent-trial design (FWER=0.0975; FMER=0.0025; MSFP=0.000625). All results are generated under the global null hypothesis at a significance level of $\alpha = 0.05$.

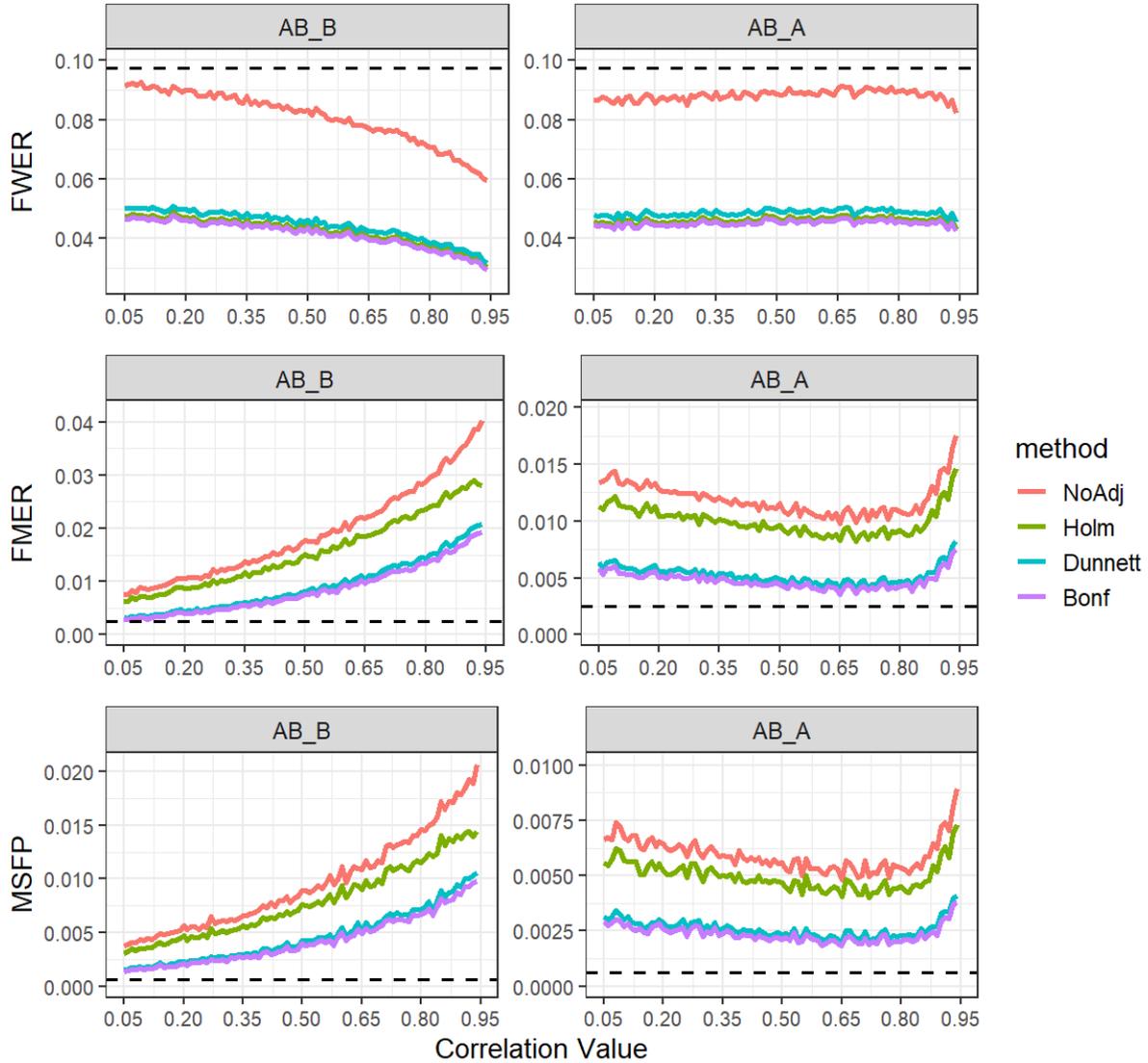

**Figure 5. False positive metrics as functions of arm correlations under four multiple testing approaches**: no adjustment (NoAdj), Holm, Dunnett's test, and Bonferroni (Bonf). The left column shows results as $\rho_{AB,B}$ varies from 0.05 to 0.95, while the right column shows results as $\rho_{AB,A}$ varies. The horizontal dashed lines represent the baseline rates under independent trials (FWER=0.0975; FMER=0.0025; MSFP=0.000625). Each panel depicts the nominal false positive rate at $\alpha = 0.05$ for a given metric and correlation scenario.

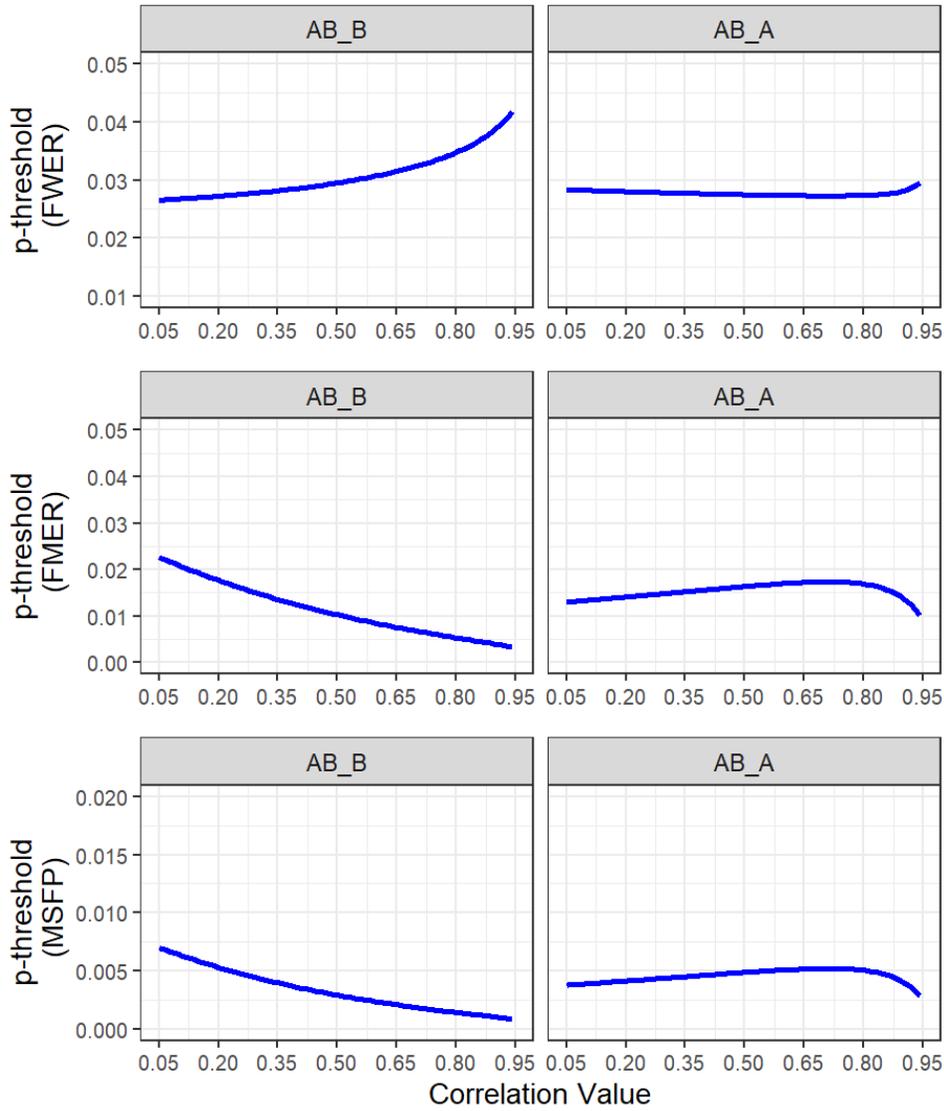

**Figure 6. P-value thresholds computed via the generalized Dunnett's procedure to control false positives at pre-specified levels**. Each row corresponds to a different false positive metric: FWER (top), FMER (middle), and MSFP (bottom). The left column shows how thresholds change as $\rho_{AB,B}$ varies from 0.05 to 0.95, while the right column shows thresholds as $\rho_{AB,A}$ varies. Target false positive rates are set at FWER=0.05, FMER=0.0025, and MSFP=0.000625, with the latter two corresponding to their respective independent-trial levels.

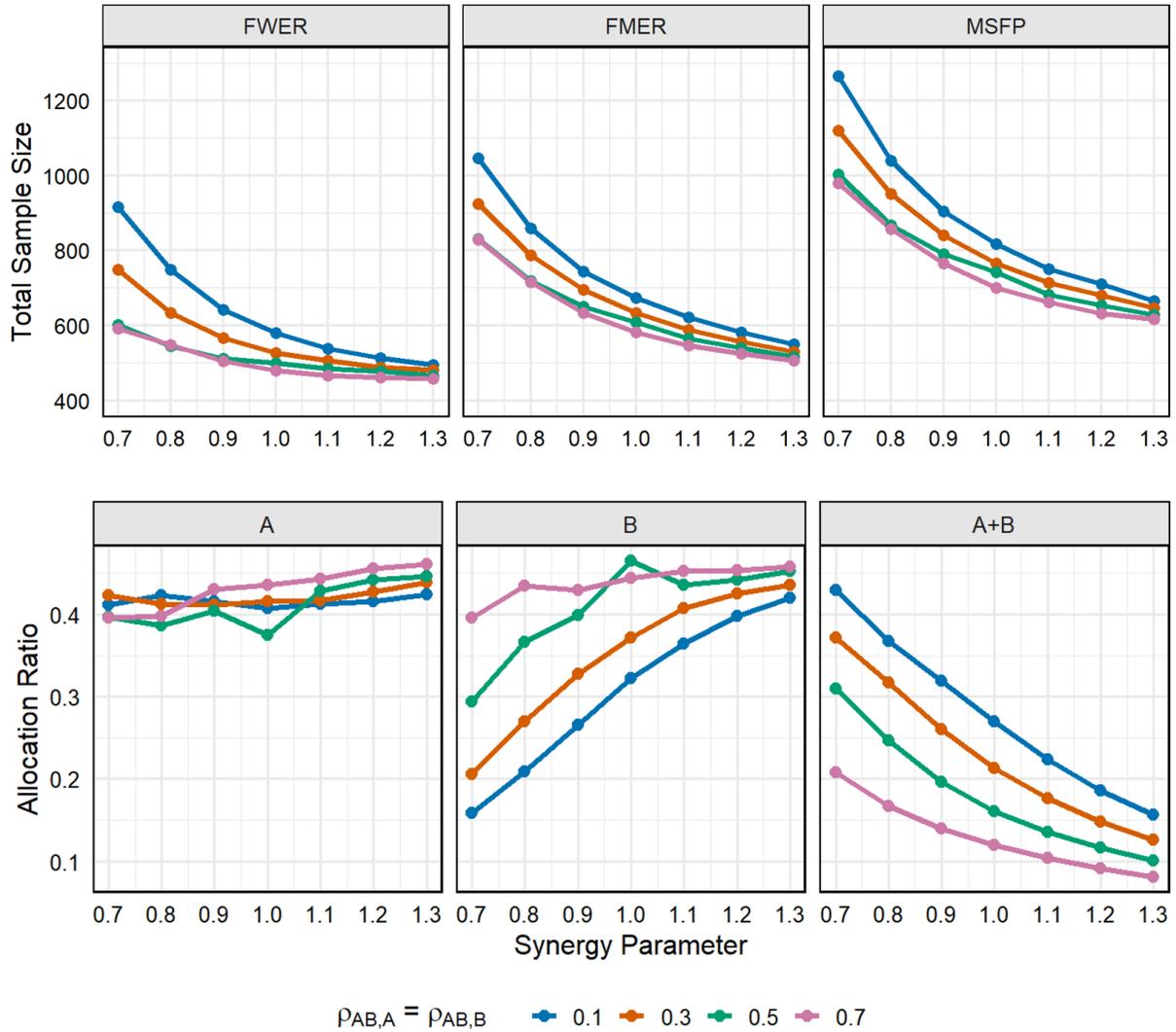

**Figure 7. Optimal allocation ratios and required sample sizes under varied synergy and arm correlation scenarios.** Each colored line represents a different correlation level ($\rho_{AB,A} = \rho_{AB,B}$) set to four values. **Top Row:** The total sample size required to achieve 80% power against the synergy parameter under three false-positive control targets (FWER = 0.05, FMER = 0.0025, and MSFP = 0.000625). **Bottom Row:** The corresponding optimal allocation ratios for the control arm $A$, the monotherapy arm $B$, and the combination arm $A + B$.

**Table 1. Trail parameter estimation and false positive control for six synthetic combination trials based on PDX preclinical data.** The p-value threshold is to control FWER = 0.05, FMER = 0.0025, and MSFP = 0.000625. To maintain consistency with the notations in Section 4, note that $\hat{\delta}_B = \hat{\delta}$ and $\hat{\delta}_{AB} = \hat{s}\hat{\delta}$. For clarity, the table omits the explicit values of $\hat{\delta}_B$ and $\hat{\delta}_{AB}$, as well as the name of the combination therapy $A + B$.

| A | B | $\hat{\rho}_{AB,A}$ | $\hat{\rho}_{AB,B}$ | $\hat{\delta}$ | $\hat{s}$ | $\hat{\rho}$ | Error metric | Error without adjustment | p-value threshold |
|---|---|---|---|---|---|---|---|---|---|
| LEE011 | everolimus | 0.227 | 0.250 | 0.329 | 2.283 | 0.461 | FWER | 0.092 | 0.027 |
| | | | | | | | FMER | 0.008 | 0.022 |
| | | | | | | | MSFP | 0.004 | 0.013 |
| LEE011 | binimetinib | 0.607 | 0.711 | 0.315 | 4.384 | 0.339 | FWER | 0.094 | 0.026 |
| | | | | | | | FMER | 0.006 | 0.030 |
| | | | | | | | MSFP | 0.003 | 0.019 |
| INC280 | trastuzumab | 0.517 | 0.318 | 0.096 | 3.663 | 0.382 | FWER | 0.094 | 0.026 |
| | | | | | | | FMER | 0.006 | 0.027 |
| | | | | | | | MSFP | 0.003 | 0.017 |
| encorafenib | binimetinib | 0.626 | 0.660 | 0.663 | 1.161 | 0.371 | FWER | 0.094 | 0.026 |
| | | | | | | | FMER | 0.006 | 0.028 |
| | | | | | | | MSFP | 0.003 | 0.017 |
| BYL719 | binimetinib | 0.510 | 0.636 | 0.067 | 7.528 | 0.494 | FWER | 0.091 | 0.027 |
| | | | | | | | FMER | 0.009 | 0.020 |
| | | | | | | | MSFP | 0.005 | 0.012 |
| BKM120 | binimetinib | 0.552 | 0.460 | 0.028 | 18.392 | 0.358 | FWER | 0.094 | 0.026 |
| | | | | | | | FMER | 0.006 | 0.029 |
| | | | | | | | MSFP | 0.003 | 0.018 |

**Table 2. Optimal allocation ratios and required sample sizes for six synthetic combination trials based on PDX preclinical data.** The power target is set to 80% and the p-value threshold is to control FWER = 0.05, FMER = 0.0025, and MSFP = 0.000625.

| A | B | $p_A^*$ | $p_B^*$ | $p_{AB}^*$ | Error metric | p-value threshold | $N^*$ |
|---|---|---|---|---|---|---|---|
| LEE011 | everolimus | 0.501 | 0.455 | 0.044 | FWER | 0.026 | 365 |
| | | | | | FMER | 0.026 | 365 |
| | | | | | MSFP | 0.006 | 443 |
| LEE011 | binimetinib | 0.491 | 0.498 | 0.011 | FWER | 0.026 | 405 |
| | | | | | FMER | 0.029 | 397 |
| | | | | | MSFP | 0.007 | 479 |
| INC280 | trastuzumab | 0.492 | 0.492 | 0.017 | FWER | 0.025 | 4746 |
| | | | | | FMER | 0.052 | 3938 |
| | | | | | MSFP | 0.013 | 4765 |
| encorafenib | binimetinib | 0.445 | 0.450 | 0.105 | FWER | 0.027 | 97 |
| | | | | | FMER | 0.013 | 114 |
| | | | | | MSFP | 0.003 | 135 |
| BYL719 | binimetinib | 0.527 | 0.463 | 0.010 | FWER | 0.026 | 9321 |
| | | | | | FMER | 0.027 | 9239 |
| | | | | | MSFP | 0.006 | 11220 |
| BKM120 | binimetinib | 0.527 | 0.462 | 0.011 | FWER | 0.025 | 52886 |
| | | | | | FMER | 0.042 | 46090 |
| | | | | | MSFP | 0.010 | 55412 |